\documentclass[11pt]{iopart}
\usepackage{iopams}
\usepackage{amsfonts}

\usepackage{cite}

\newcommand\aref[1]{appendix \ref{#1}}

\usepackage{suffix}

\usepackage[T1]{fontenc}
\usepackage[utf8]{inputenc}

\usepackage{microtype}

\usepackage[british]{babel}

\usepackage{acro}
\DeclareAcronym{lqc}{
  short={LQC},
  long={loop quantum cosmology}
}
\DeclareAcronym{lqg}{
  short={LQG},
  long={loop quantum gravity}
}
\DeclareAcronym{qft}{
  short={QFT},
  long={quantum field theory},
  long-plural-form={quantum field theories}
}
\DeclareAcronym{gft}{
  short={GFT},
  long={group field theory},
  long-plural-form={group field theories}
}
\DeclareAcronym{pg}{
  short={PG},
  long={Perelomov--Gilmore}
}
\DeclareAcronym{bg}{
  short={BG},
  long={Barut--Girardello}
}
\DeclareAcronym{flrw}{
  short={FLRW},
  long={Friedmann--Lemaître--Robertson--Walker}
}
 
\usepackage{hyperref}

\usepackage{graphicx}

\usepackage{latexsym}
\usepackage{tensor}

\DeclareMathAlphabet{\mathup}{OT1}{\familydefault}{m}{n}

\newcommand\mathsubscript[1]{\mathup{#1}}

\def\imagi{\mathup{i}}
\def\expe{\mathup{e}}
\newcommand\operatorname[1]{\mathup{#1}}

\newcommand\artanh{\operatorname{artanh}}
\newcommand\hermconjtext{\operatorname{h.c.}}

\newcommand\mathcomma{\,,}
\newcommand\mathperiod{\,.}

\newcommand\field[1]{\mathbb{#1}}

\def\dd{\mathrm{d}}
\newcommand\intmeasure[2][]{\dd^{#1}#2}
\newcommand\dfrac[2][]{\frac{\dd^{#1}}{\dd#2^{#1}}}
\WithSuffix\newcommand\dfrac*[3][]{\frac{\dd^{#1}#2}{\dd#3^{#1}}}

\newcommand\abs[1]{| #1 |}
\WithSuffix\newcommand\abs*[1]{\left\lvert #1 \right\rvert}
\newcommand\compconj[1]{\bar{#1}}
\WithSuffix\newcommand\compconj*[1]{\overline{#1}}

\newcommand\bigO[1]{O(#1)}
\WithSuffix\newcommand\bigO*[1]{O\left(#1\right)}

\newcommand\poissonbracket[2]{\{#1, #2\}}
\WithSuffix\newcommand\poissonbracket*[2]{\left\{#1, #2\right\}}
\newcommand\commutator[2]{[#1,#2]}
\WithSuffix\newcommand\commutator*[2]{\left[#1,#2\right]}
\newcommand\op[1]{\hat{#1}}

\newcommand\interaction{\mathsubscript{I}}
\newcommand\timeordering{\operatorname{T}}
\newcommand\critical{\mathsubscript{c}}
\newcommand\effective{\mathsubscript{eff}}
\newcommand\planck{\mathsubscript{P}}

\newcommand\liebracket[2]{[#1,#2]}
\WithSuffix\newcommand\liebracket*[2]{\left[#1,#2\right]}
\newcommand\liegroup[1]{\ensuremath{#1}}

\newcommand\liealg[1]{\ensuremath{#1}}

\newcommand\intertwiner{\mathcal{I}}

\newcommand\ket[1]{| #1 \rangle}
\WithSuffix\newcommand\ket*[1]{\left| #1 \right\rangle}
\newcommand\bra[1]{\langle #1 |}

\newcommand\expval[2][]{\langle #2 \rangle_{#1}}
\WithSuffix\newcommand\expval*[2][]{\left\langle #2 \right\rangle_{#1}}

\newcommand\covariance[2]{G(#1,#2)}
\newcommand\hermconj[1]{#1^\dagger}

\eqnobysec
\begin{document}

\title{Generalised effective cosmology from group field theory}

\author{Steffen Gielen, Axel Polaczek}
\address{School of Mathematics and Statistics,
  University of Sheffield,
  \\
  Hicks Building,
  Hounsfield Road,
  Sheffield S3 7RH,
  United Kingdom
  \\[.5em]
  School of Mathematical Sciences,
  University of Nottingham,
  \\
  University Park,
  Nottingham NG7 2RD,
  United Kingdom
}
\ead{s.c.gielen@sheffield.ac.uk, apolaczek1@sheffield.ac.uk}

\begin{abstract}
We extend various recent results regarding the derivation of effective cosmological Friedmann equations from the dynamics of group field theory (GFT). Restricting ourselves to a single GFT field mode (or fixed values of Peter--Weyl representation labels), we first consider dynamics given by a quadratic Hamiltonian, which takes the form of a squeezing operator, and then add a quartic interaction that can be seen as a toy model for interactions in full GFT. Our derivation of effective Friedmann equations does not require a mean-field approximation; we mostly follow a general approach in which these equations in fact hold for any state. The resulting cosmological equations exhibit corrections to classical Friedmann dynamics similar to those of loop quantum cosmology, leading to generic singularity resolution, but also involve further state-dependent terms. We then specify these equations to various types of coherent states, such as Fock coherent states or Perelomov--Gilmore states based on the $su(1,1)$ structure of harmonic quantum cosmology. We compute relative uncertainties of volume and energy in these states, clarifying whether they can be interpreted as semiclassical. In the interacting case, both analytical and numerical approximations are used to obtain modified cosmological dynamics. Our results clarify how effective cosmological equations derived from GFT can provide reliable approximations to the full dynamics.
\end{abstract}

\noindent{\it Keywords\/}: quantum cosmology, group field theory, coherent states

\maketitle

\section{Introduction}
\label{sec:introduction}
Spacetime singularities are among the most spectacular predictions of classical general relativity.
In the context of cosmology, the presence of an initial singularity signals a fundamental incompleteness in our understanding of the Universe, whose origin and beginning can not be explained within a classical or semiclassical treatment.
The singularity theorems of general relativity require energy conditions which may be violated in the early Universe if inflation was present, but one can nevertheless show that inflationary spacetimes are past incomplete \cite{BGV03}.
\footnote{See e.g.\ \cite{YQ18} for an analysis of conditions under which continuous or differentiable extensions of the spacetime metric beyond the past incomplete region may exist for such inflationary spacetimes.}
It is widely expected that a quantum theory of gravity is required to resolve the classical Big Bang singularity and give a fundamental basis to theoretical cosmology; a full quantum treatment of spacetime may indeed be needed to justify the assumption that the early Universe can be studied in terms of quantum fields on a classical background \cite{FLT17,TFLT19}.

On the other hand, the spacetime geometry of the early Universe was presumably very simple, describable in terms of a homogeneous isotropic background with only small perturbations.
This simplicity must ultimately be explained more fundamentally, but it also results in significant practical simplifications: at least as a first step, one can study simple homogeneous, isotropic spacetimes in quantum gravity to learn about singularity resolution.
One could hope that, as in the classical theory, this does not require understanding the nonlinear and presumably complicated dynamics of full quantum gravity.
This philosophy, systematically applied to \ac{lqg} as a candidate theory of quantum gravity, gave birth to the field of \ac{lqc} where, for models of quantum gravity coupled to a massless scalar field, the classical Big Bang singularity is indeed resolved and replaced by a `big bounce' \cite{Boj01,AS11}.
More recently \ac{lqc} has made direct contact with inflation, providing a quantum-gravitational extension of the usual semiclassical framework \cite{AAN12}.
The precise relation between \ac{lqg} and \ac{lqc} has been the focus of much work in recent years \cite{AC15,ADLP18,DLP19}.

The origin of singularity resolution in \ac{lqc} is the fundamental discreteness of the theory, manifest in discrete spectra for areas and volumes with a gap away from zero \cite{Rov04,BH11}.
This key feature of the \ac{lqg} kinematics is shared by its reformulation in terms of \ac{gft} \cite{Ori17}.
\ac{gft} interprets the {\em quanta of spacetime} seen in \ac{lqg} as excitations of a quantum field (also then a quantum field {\em of}, not {\em on} spacetime), while the dynamics of a \ac{gft} are generally defined such that its perturbative expansion corresponds to a sum over {\em spin foams}, or discrete spacetime histories of \ac{lqg} states \cite{RR01}.
The continuum limit of this sum, needed to obtain continuum  quantum geometry, is to be taken in a way similar to matrix and tensor models \cite{Ori12}.
Given the very close relation of \ac{gft} to \ac{lqg}, it is natural to ask whether cosmological models of \ac{gft} dynamics can resolve the classical Big Bang singularity in a way similar to what is seen in \ac{lqc}.

The key idea that led to the derivation of cosmological models in the \ac{gft} approach was that a spatially homogeneous quantum geometry could be understood as a type of Bose--Einstein condensate in \ac{gft} \cite{GOS14,GS16,Ori17a,PS19}.
As in other quantum field theories, such a condensate can be understood as a nonperturbative vacuum of the theory, characterised by a common quantum state for a very large number of quanta with respect to the original Fock vacuum.
The idea that spacetime could be a kind of Bose--Einstein condensate of geometric quanta had been formulated in other approaches before \cite{Hu05,KS12}, but the quantum field theory framework of \ac{gft} allows studying such a condensate with relatively conventional methods, adapted to a background-independent quantum gravity context.
In the simplest (mean-field) approximation, the equation of motion of the condensate mean field is the analogue of the usual Gross--Pitaevskii equation in condensed matter physics.
From a solution to this equation of motion, one can compute geometric observables such as the total volume of the condensate.
Dynamics are introduced, just as in \ac{lqc} and many other models of quantum cosmology \cite{BI75}, by coupling a relational matter clock, given by a free massless scalar field.
Concretely, one extracts equations for the relational volume observable $V(\phi)$, the three-dimensional volume of a condensate given a particular value of the relational clock field, and its derivatives.
These are then interpreted as effective Friedmann equations derived from the \ac{gft} condensate dynamics.
These steps were first fully implemented in \cite{OSW16,OSW17} where it was shown how such effective Friedmann equations, for a wide class of \ac{gft} models and under various simplifying assumptions, are consistent with the classical Friedmann equations at large volume while showing a bouncing behaviour at high densities very similar to the one in \ac{lqc}.
In particular, these effective Friedmann equations can reproduce the preferred `improved dynamics' form of \ac{lqc} \cite{APS06} whose derivation from Hamiltonian formulations of \ac{lqg} is a largely outstanding challenge \cite{DLP19}.

These very promising results for effective cosmological dynamics from \ac{gft} relied on assuming the emergence of a condensate regime in which the mean-field approximation is valid and \ac{gft} interactions are subdominant with respect to the quadratic (kinetic) term.
Including interactions leads to interesting modifications to effective cosmological equations, which can serve as a starting point for \ac{gft} phenomenology \cite{CPS16,PS17}.
To better understand the dependence of \ac{gft} cosmology on a mean-field approximation, a simple toy model was then studied in \cite{AGW18}.
In this model, only a single mode of a \ac{gft} field is excited and a {\em squeezing} Hamiltonian generates evolution in relational matter time $\phi$, so that a squeezed state emerges dynamically even from the Fock vacuum without assuming a mean-field regime.
The general features of a bounce at high densities and agreement with classical cosmology at large volume could be reproduced in this simpler setting.
The choice of Hamiltonian was rather ad-hoc, motivated by agreement with classical cosmology and the properties of squeezed states.
It was then shown in \cite{WEw19} that a Legendre transformation of the free \ac{gft} action, taking again the matter clock $\phi$ to define time, leads essentially to a squeezing Hamiltonian, the latter representing the dynamics of an `upside-down' harmonic oscillator with negative quadratic potential.
The results of \cite{WEw19} hence explained the agreement between the effective cosmological dynamics of a squeezing Hamiltonian and those of previous results for \ac{gft} condensates.
A different argument explaining the close connection of \ac{gft} cosmology to \ac{lqc} was given in \cite{BBC19} where it was argued that the canonical \ac{lqc} framework and the field-theoretic (bosonic) \ac{gft} cosmology can be seen as different realisations of the same $\liealg{su}(1,1)$ algebra.

The aim of this paper is to extend many of these recent results to further clarify the connection between fundamental \ac{gft} dynamics and effective cosmological equations.
Rather than using mean-field approximations, we derive general dynamical equations for operators and therefore expectation values of these operators; we work similarly to the `toy model' analysis of \cite{AGW18} but consider a more general form of the \ac{gft} dynamics.
In particular, we will go beyond quadratic Hamiltonians and add simple interactions, allowing us to connect to results such as those of \cite{CPS16} for \ac{gft} interactions which were obtained in a mean-field approximation.
We will also extend the algebraic viewpoint of \cite{BBC19} to the case of an interacting Hamiltonian.
An issue in the derivation of effective dynamical equations that we will encounter is that, in the general case, these equations involve additional expectation values or higher moments that are not directly identifiable with the variables of a classical \ac{flrw} cosmology (which is fully characterised by the scale factor or volume and the energy density in the massless scalar field).
In other words, dynamical equations for quantum expectation values require knowledge of additional initial conditions compared to a classical cosmology.
This is of course the usual situation for effective equations for quantum systems.
(See e.g.~\cite{BHT11} for a systematic discussion.)
Choosing a specific class of initial conditions, for instance by focussing on a class of coherent states, then simplifies these equations.
We will illustrate this trade-off between obtaining effective equations that are as general as possible but include a dependence on additional variables, and more specific equations in which a particular choice of state (or family of states) allows deriving equations with more direct (semiclassical) physical interpretation.
As concrete examples, we will discuss Fock coherent states which have been studied in most of the existing literature on \ac{gft} cosmology, but also coherent states based on the $\liealg{su}(1,1)$ algebra satisfied by basic \ac{gft} operators (in particular the well-known \ac{pg} coherent states).

We show a key property of simple Fock coherent states, which is that relative uncertainties of quantities like volume and energy can be made arbitrarily small at late times, so that these states are as semiclassical as desired.
This gives further justification to their interpretation as semiclassical macroscopic geometries \cite{GOS14}.
\acl{pg} states that can be thought of as elements of a \ac{gft}-like Fock space do not admit such a semiclassical interpretation and are therefore disfavoured for \ac{gft} cosmology.
 
\section{Group field theory cosmology}
\label{sec:gft_cosmo}
In this section we summarise previous work on the derivation of effective cosmological equations from group field theory, in particular the previous papers we are building on in this work.
For more details and background we point to \cite{GOS14,GS16,Ori17a,PS19}.

\subsection{The group field theory approach to quantum gravity}
\Acfp{gft} are a nonperturbative approach to quantum gravity, aiming to extend the successes of matrix and tensor models to a theory of quantum geometry in higher (in particular four) dimensions by incorporating the kinematical and dynamical structure of loop quantum gravity and spin foams \cite{Ori17,Ori12,Ori06,Fre05,Kra11}.

Concretely, rather than being based on a matrix or tensor with a number of discrete indices of finite range, a \ac{gft} is defined in terms of a field $\varphi$ depending on a number of continuous variables taking values in a Lie group.
In this sense, the purely combinatorial structure of matrix models is enriched by additional group-valued degrees of freedom, interpreted as parallel transports akin to the fundamental variables in lattice gauge theory.
Nevertheless, the main ideas are similar; a \ac{gft} perturbative expansion should generate a sum over quantum geometries and admit a consistent continuum limit.

The prototype for \ac{gft} as an approach to quantum gravity is the {\em Boulatov model} in three dimensions \cite{Bou92}.
One defines a real field
\begin{equation}
  \varphi:
  G^3 \rightarrow \field{R}
  \mathcomma
  \quad
  \varphi(g_1,g_2,g_3)
  = \varphi(g_2,g_3,g_1)
  = \varphi(g_3,g_1,g_2)
\end{equation}
with an action
\begin{equation}
\eqalign{
  S[\varphi]
  =
&
  \frac{1}{2}\int
  \intmeasure[3]{g} \,
  \varphi(g_1,g_2,g_3)
  \varphi(g_1,g_2,g_3)
\\
&
  -\frac{\lambda}{4!}\int
  \intmeasure[6]{g} \,
  \varphi(g_1,g_2,g_3)\varphi(g_1,g_4,g_5)\varphi(g_2,g_5,g_6)\varphi(g_3,g_6,g_4)
  \mathcomma
}
\end{equation}
where $G$ is a Lie group and $\intmeasure{g}$ is the Haar measure on $G$.
Notice the `nonlocal' combinatorial pairing of field arguments in the interaction term which is the generalisation of trace invariants such as ${\rm tr}M^n$ in the case of a matrix model.
One can then show that, for $G=\liegroup{SU}(2)$, the \ac{gft} partition function admits a perturbative expansion of the form
\begin{equation}
  \label{eq:gft_boulatov_exp}
  \int \mathcal{D}\varphi\;e^{-S[\varphi]}=\sum_{C}
  \lambda^{N_{T}(C)}\sum_{\{j_l\}\in{\rm Irrep}}\prod_{l\in C} (2j_l+1)\sum_{T\in
  C}\left\{ \matrix{
    j_{T_1} & j_{T_2} & j_{T_3}
    \cr
    j_{T_4} & j_{T_5} & j_{T_6}
  }\right\}
\end{equation}
where the first sum is over all oriented three-dimensional simplicial complexes $C$, $N_T(C)$ is the number of tetrahedra in $C$, $j_l$ is an irreducible representation of ${\liegroup{SU}(2)}$ assigned to each link $l\in C$, and $\{\cdot\}$ is the Wigner $6j$-symbol associated to a tetrahedron $T\in C$ (involving its six links).
Up to the factor $\lambda^{N_{T}(C)}$, each complex $C$ appearing in \eref{eq:gft_boulatov_exp} is weighted by its {\em Ponzano--Regge state sum} \cite{PR68}, a possible definition of discrete three-dimensional quantum gravity (see e.g.~\cite{BN09}).
In this sense, the perturbative expansion of the Boulatov model generates all possible triangulations (including all topologies) each weighted by a partition function for quantum gravity on this triangulation.
This expansion is highly divergent without further regularisation \cite{BN09}.
The \ac{gft} programme aims to extend \eref{eq:gft_boulatov_exp} to more complicated models, in particular candidates for quantum gravity in four dimensions, where the Ponzano--Regge state sum is replaced by a general {\em spin foam amplitude} \cite{Rov04}: the amplitudes of the Barrett--Crane model \cite{BC98} can be obtained as Feynman amplitudes of a \ac{gft} defined on the three-sphere $S^3=\liegroup{SO}(4)/\liegroup{SO}(3)$ \cite{PFKR00} and this extends to a one-to-one correspondence between general spin foam amplitudes and their realisations as the perturbative expansion of a \ac{gft} \cite{RR01}.
This correspondence extends from spin foam models for Euclidean quantum gravity to models with Lorentzian signature such as \cite{BC00} which can be defined through a \ac{gft} on a noncompact group such as $\liegroup{SO}(3,1)$ (see e.g.~\cite{LO07}).
In this sense one could say that a \ac{gft} defines a completion of the spin foam programme in that it not only generates spin foam amplitudes for quantum gravity on a given discretisation, but also the weights in a sum over discretisations.

\subsection{Cosmology from group field theory}
For general \ac{gft} models for quantum gravity, it is difficult to make sense of a perturbative expansion of the form \eref{eq:gft_boulatov_exp}.
The number of terms quickly grows out of control as the number of building blocks is increased and there is no obvious physical meaning to truncating such an expansion to the first few terms, i.e. to discretisations with very few building blocks.
\Eref{eq:gft_boulatov_exp} is really an expansion around a `no-space' vacuum in which no geometry is present at all.

However, there is more to quantum field theory than a perturbative expansion
around vanishing field value: interacting field theories often exhibit phase transitions to a {\em condensate} characterised by a nonvanishing field expectation value.
With respect to the original Fock vacuum in which the field vanishes, a condensate has a very large number of quanta all characterised by a single quantum state (the `condensate wavefunction').
This is a quantum state of high symmetry and quantum coherence.
The key idea of {\em \ac{gft} condensates} is that such a configuration in \ac{gft} is a candidate for a macroscopic, nearly homogeneous Universe, and hence  a starting point for effective cosmology.
We refer the reader to \cite{GOS14} for details and arguments for this geometric interpretation.

We generally define a \ac{gft} field for a four-dimensional quantum gravity model coupled to scalar matter by
\begin{equation}
  \varphi:
  G^4\times \field{R} \rightarrow \field{K}
  \mathcomma
  \quad\varphi(g_1,\ldots,g_4,\phi)=\varphi(g_1h,\ldots,g_4h,\phi)\;\forall h\in G
  \mathperiod
\end{equation}
where $G$ is the gauge group of gravity (often assumed to be $\liegroup{SU}(2)$) and $\field{K}$ is either the real or complex numbers.
The action takes the general form
\begin{equation}
  S
  =
  \int
  \intmeasure[4]{g}\,
  \intmeasure{\phi}\,
  \bar\varphi(g_I,\phi)\,\mathcal{K}\varphi(g_I,\phi)
  + \mathcal{V}[\varphi]
\end{equation}
where for a real field $\bar\varphi=\varphi$ (and, to obtain the usual normalisation of a kinetic term, one has to also rescale the field), $\mathcal{K}$ is a kinetic operator which in general contains derivatives with respect to all arguments, and all terms that  are higher order than quadratic are part of $\mathcal{V}[\varphi]$.
In general, $\mathcal{V}[\varphi]$ is also nonlocal in a way similar to the Boulatov \ac{gft} defined above.
In its Feynman expansion, such a field theory will generate graphs whose edges are labelled by $g_I$ (interpreted as parallel transports of a $G$-connection) and whose vertices are labelled by $\phi$ interpreted as the values of a matter scalar field.

If we denote the expectation value of the field operator by
\begin{equation}
  \expval{\op{\varphi}(g_I, \phi)}
  =
  \sigma(g_I,\phi)
  \mathcomma
\end{equation}
a condensate phase is then characterised by a nonvanishing $\sigma(g_I,\phi)$.

The {\em mean-field approximation} which is used in most work on \ac{gft} cosmology so far requires that the mean field $\sigma(g_I,\phi)$, for a \ac{gft} with complex field, satisfies the classical \ac{gft} equation of motion
\begin{equation}
  \label{eq:gft_cosmo_meanfield_eom}
  \mathcal{K}\sigma(g_I,\phi)
  +
  \frac{
    \delta\mathcal{V}[\sigma]
  }{
    \delta\bar\sigma(g_I,\phi)
  }
  =
  0
  \mathcomma
\end{equation}
the analogue of the Gross--Pitaevskii equation for the condensate wavefunction in condensed matter physics.
From a solution $\sigma(g_I,\phi)$ to the equation of motion, one can then extract an observable corresponding to the total condensate volume as a function of the matter field `clock',
\begin{equation}
  \label{eq:gft_cosmo_gft_volumeop}
  \expval{\op{V}(\phi)}
  \equiv
  \int
  \intmeasure[4]{g}\,
  \intmeasure[4]{g'}\,
  V(g_I,g'_I)
  \bar\sigma(g_I,\phi)
  \sigma(g'_I,\phi)
  \mathcomma
\end{equation}
where $V(g_I,g'_I)$ are matrix elements of the \ac{gft} volume operator between `single-particle' states $\ket{g_I}$ and $\ket{g'_I}$; such an operator can be defined from the action of a volume operator in \ac{lqg} on open spin networks with a single vertex and four links.
Dynamical equations satisfied by $\expval{\op{V}(\phi)}$ and its derivatives with respect to $\phi$ are then interpreted as effective cosmological (Friedmann) equations for the three-volume of (a patch of) the Universe, derived directly from a prescription for the microscopic dynamics of a \ac{gft}.

The most concrete derivation of this type, for models of quantum gravity coupled to massless scalar matter, was given in \cite{OSW16}.
First the nonlinear, nonlocal equation of motion \eref{eq:gft_cosmo_meanfield_eom} was simplified by making an `isotropic' ansatz
\begin{equation}
  \sigma(g_I,\phi)
  =
  \sum_{j\in{\rm Irrep}}
  \sigma_j(\phi){\bf D}^j(g_I)
  \mathcomma
\end{equation}
where the \ac{gft} gauge group is taken to be $\liegroup{SU}(2)$ and ${\bf D}^j(g_I)$ is a fixed convolution of four Wigner $D$-matrices for the irreducible representation $j$, encoding the `shape' of the \ac{gft} building blocks.
(${\bf D}^j(g_I)$ requires a choice of intertwiner $j\otimes j\otimes j\otimes j\rightarrow {\bf 0}$; in \cite{OSW16} this is taken to be the intertwiner with maximum eigenvalue for the volume, see \eref{eq:gft_volume_expv}.)

Assuming a quintic potential as is done for many spin foam models related to \ac{lqg}, this reduces \eref{eq:gft_cosmo_meanfield_eom} to a decoupled form
\begin{equation}
  \label{eq:gft_cosmo_isotropic_eom}
  A_j
  \partial_\phi^2\sigma_j(\phi)
  - B_j\sigma_j(\phi)
  + w_j\bar\sigma_j(\phi)^4
  =
  0
  \mathcomma
\end{equation}
where $A_j$, $B_j$ and $w_j$ are determined by the couplings in the \ac{gft} action.
Since the volume operator is diagonal when written in terms of $\liegroup{SU}(2)$ representations, the volume of a condensate in such a state is given by
\begin{equation}
  \expval{
    \op{V}(\phi)
  }
  =
  \sum_{j\in{\rm Irrep}}
  V_j\,|\sigma_j(\phi)|^2
\label{eq:gft_volume_expv}
\end{equation}
where $V_j$ is the volume eigenvalue assigned to the spin $j$ (which in general depends on the intertwiner used to define ${\bf D}^j(g_I)$).
In a regime in which interactions can be neglected, and assuming that the ratios $B_j/A_j$ take a positive maximum for some $j=j_0$, it is easy to see that for almost any solution to \eref{eq:gft_cosmo_isotropic_eom}\footnote{The only cases for which $V(\phi)$ does not have the given asymptotics are solutions for which one only uses the exponentially growing or the exponentially decaying solution to \eref{eq:gft_cosmo_isotropic_eom} for $j=j_0$.} the volume $V(\phi) \equiv \expval{\op{V}(\phi)}$ satisfies
\begin{equation}
  V(\phi)
  \stackrel{\phi\rightarrow -\infty}{\sim}
  c_1 \exp\left(-2\sqrt{\frac{B_{j_0}}{A_{j_0}}}\phi\right)
  \mathcomma
  \quad
  V(\phi)
  \stackrel{\phi\rightarrow +\infty}{\sim}
  c_2 \exp\left(2\sqrt{\frac{B_{j_0}}{A_{j_0}}}\phi\right)
\end{equation}
for some constants $c_1$ and $c_2$ \cite{Gie16}.
Moreover, $V(\phi)$ can only ever reach zero for very special initial conditions (although this case becomes generic if the \ac{gft} field is taken to be real-valued \cite{PS17}).
With the identification $\frac{B_{j_0}}{A_{j_0}}=:3\pi G$, this corresponds to a bounce solution interpolating between the expanding and contracting solutions to the {\em classical} Friedmann equations for a flat \ac{flrw} Universe filled with a massless scalar field, $V(\phi)=V_0\exp(\pm\sqrt{12 \pi G}\phi)$.
Similar conclusions apply if one considers condensates only formed by a single $j$ component, again denoted by $j_0$.
In the latter case, one can show that the volume satisfies an effective Friedmann equation \cite{OSW16}
\begin{equation}
  \label{eq:gft_cosmo_friedmann}
  \left(
    \frac{V'(\phi)}{V(\phi)}
  \right)^2
  =
  12\pi G
  \left(
    1-\frac{\rho(\phi)}{\rho_\critical}
  \right)
  + \frac{4V_{j_0}E}{V(\phi)}
\end{equation}
where $\rho=\pi_\phi^2/(2V^2)$, with $\pi_\phi$ the conserved momentum conjugate to $\phi$, is the energy density of the massless scalar field, and $\rho_\critical$ is a maximal (critical) energy density similar to the one found in \ac{lqc} \cite{Boj01,AS11} (and we have again set $\frac{B_{j_0}}{A_{j_0}}=:3\pi G$).
The last term, involving a conserved quantity $E$ (`\ac{gft} energy'), represents a slight modification with respect to the usual \ac{lqc} effective dynamics.
Again, clearly at large volumes or late times such effective dynamics reduce to the classical Friedmann equation $(V'/V)^2=12\pi G$.

In this article, we strengthen the foundations of these past results.
We aim to obtain effective cosmological dynamics from \ac{gft} without several assumptions that were necessary to obtain \eref{eq:gft_cosmo_friedmann}, namely: the validity of a mean-field regime in which one solves equations for the mean field; restriction of the effective equations to simple expectation values such as $\expval{\op{V}(\phi)}$ without taking into account fluctuations around these expectation values; neglecting \ac{gft} interactions by effectively setting $w_j=0$.
\footnote{The work of \cite{CPS16} included \ac{gft} interactions into the analysis, leading to additional terms on the right-hand side of \eref{eq:gft_cosmo_friedmann}, while working in a mean-field regime.}
Indirectly, the results outlined so far also assumed a given Fock space structure used to define a \ac{gft} volume operator, which has not been derived from the canonical analysis of a \ac{gft} action.

\subsection{Toy model for group field cosmology, and a Hamiltonian for \ac{gft}}
A first step towards deriving effective cosmological dynamics from \ac{gft} outside of a mean-field regime was taken in \cite{AGW18}.
One motivation for this work was to develop a toy model in which some of the successes of \ac{gft} cosmology could be obtained in a simpler setting, but there was also a new technical assumption: the massless scalar field $\phi$ was proposed as a (relational) time variable, with a Hamiltonian generating evolution with respect to this clock.
That is, the idea was to define a {\em deparametrised} setting in which some degrees of freedom serve as coordinates parametrising the remaining ones, a strategy widely employed in canonical quantum gravity \cite{BK95,DGKL10}.\footnote{This strategy can be extended to a \ac{gft} coupled to four massless scalar fields serving as relational coordinates for both time and space \cite{Gie18}.}
This approach was different from previous work on \ac{gft} cosmology in which the fundamental \ac{gft} formalism treated all arguments of the field on the same footing.
The Hamiltonian itself was chosen so as to reproduce the correct cosmological dynamics at large volume.

Classical \ac{flrw} cosmology can be defined by a volume variable $V(\phi)$ and conjugate momentum $p_V(\phi)$ subject to a Hamiltonian
\begin{equation}
  \label{eq:gft_cosmo_dilat}
  \mathcal{H}
  =
  \sqrt{12 \pi G}\,Vp_V
  \mathcomma
\end{equation}
generating a dilatation as its time evolution, i.e.~the exponential solutions in $\phi$ mentioned above.
In \cite{AGW18} it was then observed that, for a Fock space generated by annihilation operators $\op{A}^i$ and creation operators $\hermconj{\op{A}}_j$ (here $i,j$ run from 1 to 5) with algebra
\begin{equation}
  \commutator{
    \op{A}^i
  }{
    \hermconj{\op{A}}_j
  }
  =
  \delta^i_j
  \mathcomma
\end{equation}
a discrete analogue of the dilatation operator is given by a {\em squeezing Hamiltonian}
\begin{equation}
  \label{eq:gft_cosmo_squeezing}
  \op{\mathcal{H}}
  =
  \frac{\imagi}{2}
  \lambda
  \left(
    \hermconj{\op{A}}_i \hermconj{\op{A}}_j \epsilon^{ij}
    - \op{A}^i \op{A}^j \epsilon_{ij}
  \right)
  \mathcomma
\end{equation}
where $\epsilon^{ij}$ is an appropriate symmetric tensor.
Indeed, for a volume operator taken to be the multiple of the number operator
\begin{equation}
  \label{eq:gft_cosmo_volume_via_number_op}
  \op{V}
  =
  v_0 \op{N}
  :=
  v_0 \hermconj{\op{A}}_i \op{A}^i
\end{equation}
one can show that, for suitable states characterised by the eigenvalues of $\op{V}$, $\op{\mathcal{H}}$ acts as
\begin{equation}
  (\op{\mathcal{H} }\Psi)(V)
  \stackrel{v_0\rightarrow 0}{\rightarrow}
  - \imagi\lambda\left(V\frac{\partial}{\partial V}
  + \frac{\partial}{\partial V}V\right)\Psi(V)
  \mathperiod
\end{equation}
Thus, with the identification $\lambda:=\sqrt{3\pi G}$ the continuum limit of  squeezing \eref{eq:gft_cosmo_squeezing} is compatible with the classical dilatation Hamiltonian \eref{eq:gft_cosmo_dilat}.
The picture of a Fock space of `quanta of geometry' in which each quantum carries a given volume mimics the Fock space structure of \ac{gft}, with the simplification that here each quantum comes with a fixed $v_0$ rather than a general state-dependent volume as in \eref{eq:gft_cosmo_gft_volumeop}.

Given that the Hamiltonian is quadratic, expressions for the time evolution of observables of interest can be computed analytically.
One finds that
\begin{equation}
  \label{eq:gft_cosmo_toymodel_n}
  \expval{\op{N}(\phi)}
  =
  - \frac{5}{2}
  +
  \left(
    N_0 + \frac{5}{2}
  \right)
  \cosh(2\lambda\phi)
  + Q \sinh(2\lambda\phi)
\end{equation}
with
\begin{equation}
  N_0
  :=
  \left.
    \expval{\op{N}}
  \right|_{\phi=0}
  \mathcomma
  \quad
  Q
  :=
  \frac{1}{2}
  \left.
    \left(
      \epsilon^{ij} \expval{\hermconj{\op{A}}_i \hermconj{\op{A}}_j}
      + \epsilon_{ij} \expval{\op{A}^i \op{A}^j}
    \right)
  \right|_{\phi=0}
\end{equation}
(computed equivalently in the Schrödinger or Heisenberg picture).
At late or early times $\phi\rightarrow\pm\infty$ (and with $\lambda:=\sqrt{3\pi G}$), the expectation value $V(\phi)\equiv\expval{\op{V}(\phi)}$ then again reproduces the classical solution $V(\phi)=V_0\exp(\pm\sqrt{12\pi G}\phi)$\,.
Moreover, one can show that only special initial conditions (such as choosing the Fock vacuum as initial state) lead to a solution that ever encounters a singularity where $V(\phi_0)=0$ for some $\phi_0$.
Generic solutions avoid the classical singularity and describe a bounce connecting the classical expanding and contracting branches.

The quantity $Q$ cannot be directly interpreted in terms of the volume or energy density of the corresponding classical cosmology; its presence leads to an asymmetry in the solution \eref{eq:gft_cosmo_toymodel_n}. For simplicity, the further analysis of \cite{AGW18} only considered the case $Q=0$, for which one obtains the effective Friedmann equation
\begin{equation}
  \left(
    \frac{V'(\phi)}{V(\phi)}
  \right)^2
  =
  4\lambda^2
  \left(
    1
    + \frac{5v_0}{V(\phi)}
    - \frac{N_0(N_0+5)v_0^2}{V(\phi)^2}
  \right)
  \mathperiod
\end{equation}
The similarity to the effective Friedmann equation \eref{eq:gft_cosmo_friedmann} of \ac{gft} in the mean-field setting is apparent.
In this sense, the toy model based on a squeezing Hamiltonian \eref{eq:gft_cosmo_squeezing} already reproduced several previous results in \ac{gft} cosmology.

One reason for this close connection was uncovered in \cite{WEw19,GPW19} where, taking again a deparametrised viewpoint, a Hamiltonian formalism was derived from a Legendre transformation of the full (free) \ac{gft} action in which the `matter' argument $\phi$ of the \ac{gft} field is taken as a time coordinate.
The starting point is an action for real \ac{gft} fields of the form
\begin{equation}
  \label{eq:gft_cosmo_gft_hamiltonian_action}
  S
  =
  \frac{1}{2}
  \int
  \intmeasure{\phi}
  \sum_{\vec{\jmath},\vec{m},\iota}
  \varphi^{\vec{\jmath},\iota}_{-\vec{m}}(\phi)
  \left[
    \mathcal{K}_{\vec{\jmath},\vec{m},\iota}^{(0)}
    + \mathcal{K}_{\vec{\jmath},\vec{m},\iota}^{(2)}\partial_\phi^2
  \right]
  \varphi^{\vec{\jmath},\iota}_{\vec{m}}(\phi)
  + \mathcal{V}[\varphi]
  \mathcomma
\end{equation}
where the field $\varphi(g_I,\phi)$ has been decomposed into Peter--Weyl modes according to
\begin{equation}
  \varphi(g_I,\phi)
  =
  \sum_{j_I\in{\rm Irrep}}
  \sum_{m_I,n_I=-j_I}^{j_I}
  \sum_{\iota}\varphi^{\vec{\jmath},\iota}_{\vec{m}}(\phi)\,
  \intertwiner^{\vec{\jmath},\iota}_{\vec{n}}
  \prod_{I=1}^4
  \sqrt{2j_I+1}\,
  D^{j_I}_{m_I n_I}(g_I)
\end{equation}
and $\iota$ labels a basis of intertwiners $\intertwiner$ for the representation labels $\{j_I\}$ (and the sum over $(\vec{\jmath},\vec{m},\iota)$ in \eref{eq:gft_cosmo_gft_hamiltonian_action} is a shorthand for the sums appearing in this decomposition).
For a real field, these Peter--Weyl coefficients satisfy the reality condition
\begin{equation}
  \compconj*{
    \varphi^{\vec{\jmath},\iota}_{\vec{m}}(\phi)
  }
  =
  (-1)^{\sum_I (j_I-m_I)}\varphi^{\vec{\jmath},\iota}_{-\vec{m}}(\phi)
  \mathperiod
\end{equation}
The Hamiltonian can then be written as
\begin{equation}
  \mathcal{H}
  =
  - \frac{1}{2}
  \sum_{\vec{\jmath},\vec{m},\iota}
  \left[
    \frac{
      \pi^{\vec{\jmath},\iota}_{\vec{m}}\pi^{\vec{\jmath},\iota}_{-\vec{m}}
    }{
      \mathcal{K}_{\vec{\jmath},\vec{m},\iota}^{(2)}
    }
    +
    \mathcal{K}_{\vec{\jmath},\vec{m},\iota}^{(0)}
    \varphi^{\vec{\jmath},\iota}_{\vec{m}}
    \varphi^{\vec{\jmath},\iota}_{-\vec{m}}
  \right]
  - \mathcal{V}[\varphi]
\end{equation}
whose free part corresponds to either a harmonic oscillator or an `upside down' harmonic oscillator for each mode, depending on the signs of the couplings $\mathcal{K}_{\vec{\jmath},\vec{m},\iota}^{(0)}$ and $\mathcal{K}_{\vec{\jmath},\vec{m},\iota}^{(2)}$.
Annihilation and creation operators can then be defined by
\begin{eqnarray}
  \label{eq:gft_cosmo_annihilation}
  \op{a}_{\vec{\jmath},\vec{m},\iota}
  & = &
  \frac{1}{
    \sqrt{
      2
      \abs{\mathcal{K}^{(2)}_{\vec{j}, \vec{m}, \iota}}
      \omega^{\vec{\jmath},\iota}_{\vec{m}}
    }
  }
  \left(
    \abs{\mathcal{K}^{(2)}_{\vec{j}, \vec{m}, \iota}}
    \omega^{\vec{\jmath},\iota}_{\vec{m}}
    \op\varphi^{\vec{\jmath},\iota}_{\vec{m}}
    +\imagi (-1)^{\sum_I (j_I-m_I)}\op\pi^{\vec{\jmath},\iota}_{-\vec{m}}
  \right)
  \\
  \label{eq:gft_cosmo_creation}
  \hermconj{\op{a}}_{\vec{\jmath},\vec{m},\iota}
  & = &
  \frac{1}{
    \sqrt{
      2
      \abs{\mathcal{K}^{(2)}_{\vec{j}, \vec{m}, \iota}}
      \omega^{\vec{\jmath},\iota}_{\vec{m}}
    }
  }
  \left(
    (-1)^{\sum_I (j_I-m_I)}
    \abs{\mathcal{K}^{(2)}_{\vec{j}, \vec{m}, \iota}}
    \omega^{\vec{\jmath},\iota}_{\vec{m}}
    \op\varphi^{\vec{\jmath},\iota}_{-\vec{m}}
    -\imagi \op\pi^{\vec{\jmath},\iota}_{\vec{m}}
  \right)
  \mathcomma
\end{eqnarray}
where
$
\omega^{\vec{\jmath},\iota}_{\vec{m}}
=
\sqrt{
  \abs{
    \mathcal{K}_{\vec{\jmath},\vec{m},\iota}^{(0)}
    / \mathcal{K}_{\vec{\jmath},\vec{m},\iota}^{(2)}
  }
}
$.
The free Hamiltonian is then $\op{\mathcal{H}} = \sum_{\vec{\jmath},\vec{m},\iota} \op{\mathcal{H}}_{\vec{\jmath},\vec{m},\iota}$ written as a sum of single-mode Hamiltonians $\op{\mathcal{H}}_{\vec{\jmath},\vec{m},\iota}$.
For a mode for which $\mathcal{K}_{\vec{\jmath},\vec{m},\iota}^{(0)}$ and $\mathcal{K}_{\vec{\jmath},\vec{m},\iota}^{(2)}$ have different signs the single-mode Hamiltonian is given by
\begin{equation}
  \op{\mathcal{H}}_{\vec{\jmath},\vec{m},\iota}
  =
  -\frac{1}{2}
  {\rm sgn}(\mathcal{K}_{\vec{\jmath},\vec{m},\iota}^{(0)})
  \omega_{\vec{m}}^{\vec{j}, \iota}
  \left(
    \hermconj{\op{a}}_{\vec{\jmath},\vec{m},\iota}
    \hermconj{\op{a}}_{\vec{\jmath},-\vec{m},\iota}
    + \op{a}_{\vec{\jmath},\vec{m},\iota}
    \op{a}_{\vec{\jmath},-\vec{m},\iota}
  \right)
\end{equation}
which is analogous to the squeezing operator \eref{eq:gft_cosmo_squeezing} (after redefinition by a phase $\op{A}^i \rightarrow e^{\imagi\pi/4}\op{A}^i$ and $\hermconj{\op{A}}_j \rightarrow e^{-\imagi\pi/4}\hermconj{\op{A}}_j$, \eref{eq:gft_cosmo_squeezing} becomes $\op{\mathcal{H}}=\frac{1}{2}\lambda(\hermconj{\op{A}}_i \hermconj{\op{A}}_j\epsilon^{ij}+\op{A}^i\op{A}^j\epsilon_{ij})$).
In this sense, at least for modes with magnetic indices $m_i=0$ for which there is no coupling between modes, the Hamiltonian dynamics coming from the quadratic part of the full \ac{gft} action is exactly of squeezing type.
 
\section{A toy model revisited}
\label{sec:tm}
In this section, we  study the dynamics of \ac{gft} for a single field mode, in the approximation where \ac{gft} interactions are neglected.
The observation that a squeezing operator as used in \cite{AGW18} is already the (free) \ac{gft} Hamiltonian for a mode in which all the magnetic indices are zero motivates us to revisit the model studied in \cite{AGW18}.
The restriction of this model to a single field mode can be partially motivated by results in \cite{Gie16} that suggest \ac{gft} dynamics are generically dominated by a single value for the spin $j$.
In the next section, we will add interactions and go beyond the assumption of free dynamics.

We make use of the observation of \cite{BBC19} that the fundamental operators appearing in this model (representing the Hamiltonian and volume operators) generate an $\liealg{su}(1, 1)$ algebra.
We will extend some results both of the toy model analysis \cite{AGW18} and of the algebraic discussion in \cite{BBC19}: we will discuss general algebraic expressions representing effective Friedmann equations, and then specify by choosing different classes of coherent states.
Importantly, we will compute relative uncertainties for the main physical quantities and use them as a criterion for the selection of good semiclassical states.

The Hamiltonian we consider is the one-mode squeezing Hamiltonian
\begin{equation}
  \label{eq:tm_hamiltonian}
  \op{H}
  =
  - \frac{\omega}{2}
  (
    \op{a}^2
    + \hermconj{\op{a}}{}^2
  )
  \mathperiod
\end{equation}
As in previous work the main observable of interest is the volume operator
\begin{equation}
  \label{eq:tm_n_v_relation}
  \op{V}
  =
  v_0
  \op{N}
  :=
  v_0
  \hermconj{\op{a}}
  \op{a}
  \mathcomma
\end{equation}
where $v_0$ would be the eigenvalue for the \ac{gft} volume operator for the representation (and intertwiner) chosen for the model, i.e., the volume associated to a single quantum in this mode.
We are working in a deparametrised framework in which the Hamiltonian generates time evolution with respect to scalar field time $\phi$.
The energy expectation value $\expval{\op{H}}$ thus physically represents the conjugate momentum $\pi_\phi$ of $\phi$.
We can then define an effective energy density of the matter scalar field $\phi$, at the level of expectation values, by
\begin{equation}
  \label{eq:tm_energy_density}
  \rho_\phi(\phi)
  =
  \frac{
    \expval{\op{H}}^2
  }{
    2 \expval{\op{V}(\phi)}^2
  }
  \mathcomma
\end{equation}
given the classical expression $\rho_\phi=\pi_\phi^2/(2V^2)$.
This definition extends the one used in the mean-field setting (see \eref{eq:gft_cosmo_friedmann}) which also only included expectation values of elementary operators $\op{H}$ and $\op{V}$.
Other definitions using composite operators would be possible and would differ from  \eref{eq:tm_energy_density} by higher moments such as $\expval{\op{H^2}}-\expval{\op{H}}^2$.
Notice that inverse operators such as $\op{V}(\phi)^{-2}$ are not obviously well-defined in the \ac{gft} formalism.

\subsection{\texorpdfstring{$\liealg{su}(1, 1)$}{su(1, 1)} structure of the system}
As was first pointed out for \ac{gft} cosmology in \cite{BBC19}, the operators $\op{V}$ and $\op{H}$ generate the Lie algebra $\liealg{su}(1, 1)$, which appears frequently in the context of quantum cosmology for a flat \ac{flrw} Universe filled with a free scalar field, see e.g.\ \cite{Boj07,EM12,AL19}.
The three independent quadratic products of creation and annihilation operators form a realisation of the $\liealg{su}(1,1)$ algebra.
In particular the identifications
\begin{equation}
  \label{eq:su11_bosonic_realisation}
  \op{K}_0
  =
  \frac{1}{4}
  \left(
    \hermconj{\op{a}}
    \op{a}
    +
    \op{a}
    \hermconj{\op{a}}
  \right)
  = \frac{1}{2}\op{N}+\frac{1}{4}\op{I}
  \mathcomma
  \qquad
  \op{K}_+
  =
  \frac{1}{2}
  \hermconj{\op{a}}{}^2
  \mathcomma
  \qquad
  \op{K}_-
  =
  \frac{1}{2}
  \op{a}^2
\end{equation}
(where $\op{I}$ denotes the identity)
give the $\liealg{su}(1,1)$ relations with the usual normalisation
\begin{equation}
  \commutator{
    \op{K}_0
  }{
    \op{K}_\pm
  }
  =
  \pm \op{K}_\pm
  \mathcomma
  \qquad
  \commutator{
    \op{K}_-
  }{
    \op{K}_+
  }
  =
  2 \op{K}_0
  \mathperiod
\end{equation}
The Casimir of $\liealg{su}(1, 1)$ is given by
\begin{equation}
  \label{eq:su11_casimir}
  \op{C}
  =
  (\op{K}_0)^2
  -
  \frac{1}{2}
  (
    \op{K}_+
    \op{K}_-
    +
    \op{K}_-
    \op{K}_+
  )
  \mathperiod
\end{equation}
In terms of the $\liealg{su}(1, 1)$ generators the Hamiltonian \eref{eq:tm_hamiltonian} reads
\begin{equation}
  \label{eq:tm_hamiltonian_su11}
  \op{H}
  =
  - \omega
  (
    \op{K}_+
    +
    \op{K}_-
  )
  \mathperiod
\end{equation}
As one can see from \eref{eq:su11_bosonic_realisation} the dynamics of the operator $\op{K}_0$ are intimately related to the dynamics of the volume operator $\op{V} = v_0 \op{N}$.
We consider here only the $\liealg{su}(1, 1)$ representations of the discrete ascending series in which the operator $\op{K}_0$, and hence the volume, is bounded from below.\footnote{In general, for such representations one can only say that $\expval{\op{N}}>-\frac{1}{2}$.
Below we mostly focus on Fock representations, for which $\op{N}$ is always nonnegative.}
(A more general discussion would also include other types of representations, for which there is no such lower bound. See also the comments in \cite[Sec.~4]{BBC19}.)

These representations are labelled by a real positive number $k$, the so-called Bargmann index, and satisfy
\begin{eqnarray}
  &
  \op{K}_-
  \ket{k, 0}
  =
  0
  \mathcomma
  \\
  &
  \op{K}_0
  \ket{k, m}
  =
  (k + m)
  \ket{k, m}
  \mathcomma
  \\
  &
  \op{C}
  \ket{k, m}
  =
  k (k-1)
  \ket{k, m}
  \mathcomma
\end{eqnarray}
where the states $\ket{k, m}$ are the normalised states proportional to
$(\op{K}_+)^m \ket{k, 0}$.
See \aref{app:su11} for more details.

When one inserts the realisation of the $\liealg{su}(1, 1)$ operators in terms of bosonic creation and annihilation operators \eref{eq:su11_bosonic_realisation} into the Casimir \eref{eq:su11_casimir}, one finds that the Casimir is $\op{C} = - 3/16 \op{I}$ which implies a Bargmann index of either $k=1/4$ or $k=3/4$.
These two cases respectively correspond to representations spanned by the eigenstates of the number operator with even or odd eigenvalues.
The choice $k=1/4$ appears more interesting physically since it contains the Fock vacuum (or cosmological `singularity') in which no quanta are present.
Since we are interested in studying the Fock space representations of the \ac{gft} field, we will mostly restrict the Bargmann index to these cases.

\subsection{Classes of coherent states and relative uncertainties}
\label{sec:tm_class_of_coh_stat_and_rel_uncert}
The time evolution of a system can be quite sensitive to its initial state.
In this section we discuss classes of coherent states and comment on their usefulness in the context of \ac{gft} cosmology.
The coherent states we consider are the following:
\begin{itemize}
  \item
    Fock coherent states,
  \item
    \acf{pg} coherent states of $\liealg{su}(1, 1)$,
  \item
    \acf{bg} coherent states of $\liealg{su}(1, 1)$.
\end{itemize}

The Fock coherent states correspond to the well-known coherent states of the harmonic oscillator, labelled by a complex number $\sigma$.
One possible way to define the Fock coherent states is by acting with the displacement operator on the Fock vacuum,
\begin{equation}
  \label{eq:tm_coh_state_fock}
  \ket{\sigma}
  =
  \exp\left(
    \sigma \hermconj{\op{a}}
    - \compconj{\sigma} \op{a}
  \right)
  \ket{0}
  \mathperiod
\end{equation}

The \ac{pg} coherent states are obtained by acting on the $\liealg{su}(1, 1)$ ground state $\ket{k, 0}$ with a squeezing operator
\begin{equation}
  \op{S}(\xi)
  =
  \exp\left(
    \xi
    \op{K}_+
    -
    \compconj{\xi}
    \op{K}_-
  \right)
  \mathperiod
\end{equation}
We will denote the \ac{pg} coherent states by $\ket{\zeta, k}$ and they are obtained by the following choice of squeezing parameter
\begin{equation}
  \ket{
    \zeta, k
  }
  =
  \op{S}\left(
    \frac{\zeta}{\abs{\zeta}}
    \artanh{\abs{\zeta}}
  \right)
  \ket{k, 0}
  \mathcomma
  \qquad
  \abs{\zeta} < 1
  \mathperiod
\end{equation}

The \ac{bg} coherent states will be denoted by $\ket{\chi, k}$ and are defined to be the eigenstates of the lowering operator $\op{K}_-$,
\begin{equation}
  \op{K}_-
  \ket{\chi, k}
  =
  \chi
  \ket{\chi, k}
  \mathperiod
\end{equation}

We now turn to the question of which class of coherent states should be considered in the context of \ac{gft} cosmology.
In the context of quantum cosmology a commonly studied quantity is the relative uncertainty of the volume operator.
It is argued that the magnitude of the relative uncertainty corresponds to a measure of `quantumness' of the system at some given time and it is therefore important that the theory allows for initial states which give a (comparatively) small value for the relative uncertainty at late times since then the system has become (semi-)classical.
For previous works commenting on this in the context of \ac{lqc} see, e.g., \cite{AG15} and in the context of \ac{gft} see, e.g., \cite{PS17}.
In addition one would also require the relative uncertainty of the energy (which is a constant of motion) to be very small.

We define the relative uncertainty of an operator $\op{O}$ for a given state $\ket{\psi}$ as
\begin{equation}
  r(\op{O}, \ket{\psi})
  =
  \frac{
    \bra{\psi}
    \op{O}^2
    \ket{\psi}
    -
    \bra{\psi}
    \op{O}
    \ket{\psi}^2
  }{
    \bra{\psi}
    \op{O}
    \ket{\psi}^2
  }
  \mathperiod
\end{equation}
In the following we state the relative uncertainty of the Hamiltonian and the asymptotic relative uncertainty of the volume operator at large volumes (i.e., for $\phi\rightarrow\pm\infty$), for the three classes of coherent states that we are interested in.

For the Fock coherent states one obtains for the relative uncertainty of the Hamiltonian and the asymptotic relative uncertainty of the volume operator
\begin{eqnarray}
  \label{eq:tm_rel_uncertainty_h_fock}
  r(
    \op{H},
    \ket{\sigma}
  )
  =
  \frac{
    2 (1 + 2 \abs{\sigma}^2)
  }{
    (\sigma^2 + \compconj{\sigma}^2)^2
  }
  \mathcomma
  \\
  \label{eq:tm_rel_uncertainty_v_fock}
  r(
    \op{V}(\pm \infty),
    \ket{\sigma}
  )
  =
  \frac{
    2
    \left(
      1 \mp 2 \imagi ( \sigma \pm \imagi \compconj{\sigma} )^2
    \right)
  }{
    \left(
      1 \mp \imagi ( \sigma \pm \imagi \compconj{\sigma} )^2
    \right)^2
  }
  \mathperiod
\end{eqnarray}
In principle the value of the parameter $\sigma$ is arbitrary and therefore for suitable choices of $\sigma$ the asymptotic relative uncertainty in both energy and volume becomes arbitrarily small.
These states can hence be interpreted as becoming semiclassical, consistent with arguments from \ac{gft} that suggest that such `condensates' are good candidates for effective semiclassical macroscopic geometries \cite{GOS14,GS16,Ori17a,PS19}.

For the \ac{pg} coherent states the relative uncertainties of interest are
\begin{eqnarray}
  r(
    \op{H},
    \ket{\zeta, k}
  )
  =
  \frac{1}{2k}
  \frac{
    (1 + \zeta^2)
    (1 + \compconj{\zeta}^2)
  }{
    (\zeta + \compconj{\zeta})^2
  }
  \mathcomma
  \\
  r(
    \op{V}(\pm \infty),
    \ket{\zeta, k}
  )
  =
  \frac{1}{2k}
  \mathperiod
\end{eqnarray}
The asymptotic relative uncertainty of the volume operator is independent of the parameter labelling the different \ac{pg} coherent states.
This suggests that for \ac{pg} coherent states the classical limit is reached for $k \rightarrow \infty$.
However, we saw before that if we want to consider coherent states living in a bosonic Fock representation (rather than a more general $\liealg{su}(1,1)$ representation), this restricts the values of the Bargmann index to either $k = 1/4$ or $k = 3/4$.
Thus we conclude that in the context of \ac{gft} cosmology the class of \ac{pg} coherent states do not `classicalise' at late times and hence, even though these states are naturally suggested by the  $\liealg{su}(1,1)$ structure, they do not appear to be good candidate states for effective macroscopic cosmologies in \ac{gft}.

For completeness we also state the relative uncertainties for the \ac{bg} coherent states
\begin{eqnarray}
  &r(
    \op{H},
    \ket{\chi, k}
  )
  =
  \frac{2}{
    (\chi + \compconj{\chi})^2
  }
  \left[
    k
    +
    \abs{\chi}
    \frac{
      I_{2k}(2 \abs{\chi})
    }{
      I_{2k-1}(2 \abs{\chi})
    }
  \right]
  \mathcomma
  \\
&
\eqalign{
  r(
    \op{V}(\pm \infty),
    \ket{\chi, k}
  )
  =
  &
  2
  \Big[
    - 2 \abs{\chi}^2
    I_{2k}(2 \abs{\chi})^2
    +
    (3 - 4k)
    \abs{\chi}
    I_{2k}(2\abs{\chi})
    I_{2k-1}(2\abs{\chi})
    \\
  &
    \qquad
    +
    (k \mp \imagi(\chi - \compconj{\chi}) + 2 \abs{\chi}^2)
    I_{2k-1}(2\abs{\chi})^2
  \Big]
  \\
  &
  \times
  \Big[
    2 \abs{\chi}
    I_{2k}(2 \abs{\chi})
    +
    (2 k \mp \imagi(\chi - \compconj{\chi}))
    I_{2k-1}(2 \abs{\chi})
  \Big]^{-2}
  \mathcomma
}
\end{eqnarray}
where $I_\alpha(x)$ is the modified Bessel function of the first kind.
We can use an asymptotic expansion of the modified Bessel functions to get the relative uncertainties for large values of $\abs{\chi}$,
\begin{eqnarray}
  &r(
    \op{H},
    \ket{\chi, k}
  )
  \stackrel{\abs{\chi} \rightarrow \infty}{\sim}
  \frac{
    2
    \abs{\chi}
  }{
    (\chi + \compconj{\chi})^2
  }
  \mathcomma
\\
&
  r(
    \op{V}(\pm \infty),
    \ket{\chi, k}
  )
  \stackrel{\abs{\chi} \rightarrow \infty}{\sim}
  \frac{
    2
  }{
    2 \abs{\chi} \mp \imagi (\chi - \compconj{\chi})
  }
  \mathcomma
\end{eqnarray}
which shows that the asymptotic relative uncertainties for the \ac{bg} coherent states are also arbitrarily small for large values of $\abs{\chi}$.
Hence these states would also be suitable states for \ac{gft} cosmology.
However, the ubiquitous appearance of the modified Bessel functions makes calculations with the \ac{bg} states quite cumbersome.
Below we mostly focus on Fock coherent states which are easier to calculate with.

\begin{figure}[htpb]
  \centering
  \includegraphics{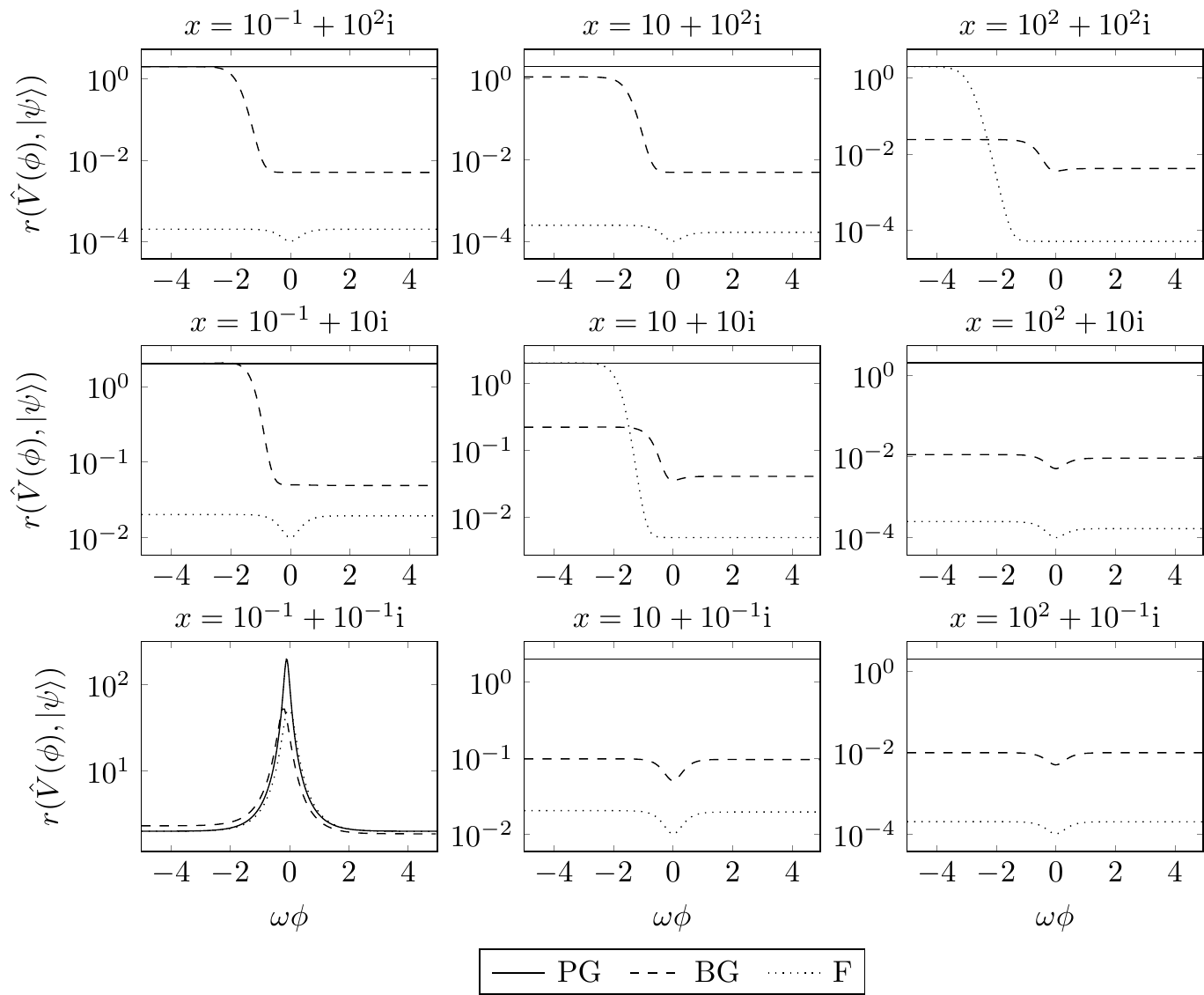}
  \caption{The relative uncertainty of the volume operator as a function of $\omega \phi$ for $k=1/4$.
    The complex parameter $x$ given in the figure is related to the parameters of the coherent states in the following manner.
    \ac{pg}:
    $\ket{\zeta, k} \equiv \ket{(x/\abs{x}) \tanh(\abs{x}),1/4}$
    \ac{bg}:
    $\ket{\chi, k} \equiv \ket{x, 1/4}$,
    Fock:
    $\ket{\sigma} \equiv \ket{x}$.
    For the bottom-left plot, $|x|\ll 1$ and these states have very small volume around $\phi=0$, which leads to the large relative uncertainties.
  }\label{fig:tm_rel_uncertainty_comp}
\end{figure}

\begin{figure}[htpb]
  \centering
  \includegraphics{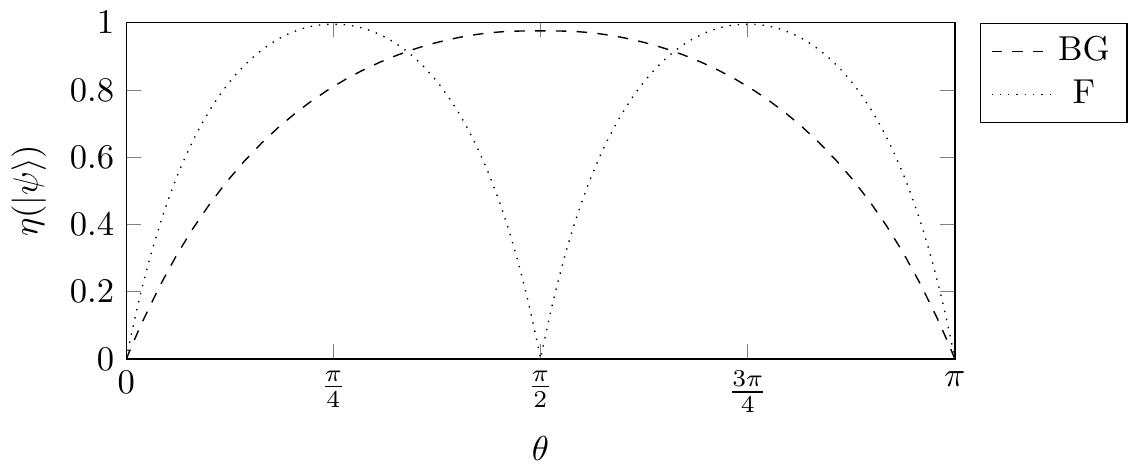}
  \caption{The asymptotic asymmetry parameter \eref{eq:tm_asymmetry_parameter} as a function of the argument of the complex coherent state parameters for $k=1/4$.
    The argument $\theta$ is related to the parameters of the coherent states in the following way.
    \ac{bg}:
    $\ket{\chi, k} \equiv \ket{100 \exp(\imagi \theta), 1/4}$,
    Fock:
    $\ket{\sigma} \equiv \ket{100 \exp(\imagi \theta)}$.
  }\label{fig:tm_asymmetry_parameter}
\end{figure}

In \fref{fig:tm_rel_uncertainty_comp} we provide an overview of the time dependence of the relative uncertainty of the various states discussed here.
One notable aspect is that the uncertainties are asymmetric with respect to time.
While the uncertainty can be asymptotically small in the future, it might have been asymptotically large in the past.
In particular, one could in general not conclude from the emergence of a semiclassical regime at late times, in which the relative uncertainties remain small, that the same was true at early times in the collapsing pre-bounce phase.
In order to quantify this asymmetry of the asymptotic relative uncertainty, we define an `asymptotic asymmetry parameter'
\begin{equation}
  \label{eq:tm_asymmetry_parameter}
  \eta(\ket{\psi})
  =
  1
  -
  \min
  \left\{
    \frac{
      r(\op{V}(+\infty), \ket{\psi})
    }{
      r(\op{V}(-\infty), \ket{\psi})
    }
    ,
    \frac{
      r(\op{V}(-\infty), \ket{\psi})
    }{
      r(\op{V}(+\infty), \ket{\psi})
    }
  \right\}
  \mathperiod
\end{equation}
The values this parameter can take lie between zero and one, where small values signify that the relative uncertainty is the same in the asymptotic past and future whereas values close to one correspond to a large past-future asymmetry.
In \fref{fig:tm_asymmetry_parameter} the asymptotic asymmetry parameter is shown for Fock and \ac{bg} coherent states as a function of the argument $\theta$ of the complex parameters characterising the state, i.e.\ $\chi = \abs{\chi} \exp(\imagi \theta)$ and $\sigma = \abs{\sigma} \exp(\imagi \theta)$.
    Note that even though the plot is done for some specific value of the absolute value of the coherent state parameters, the situation is generic.
    Only for very small absolute values ($\abs{\sigma} \ll 1, \abs{\chi} \ll 1$) is the asymmetry parameter close to one for all values of the argument $\theta$.
From this it becomes apparent that the past-future asymmetry is rather generic.
Similar questions have been discussed previously in the context of \ac{lqc} \cite{Boj07a,CP08,Boj08}; our analysis here extends them from the mean-field calculations in \cite{PS17} to broader classes of states of interest for \ac{gft} cosmology.

\subsection{Effective Friedmann equations}
In order to derive the cosmological implications of the model, we derive in this section an effective Friedmann equation
\begin{equation}
  \label{eq:tm_eff_fried_formal}
  \left(
    \frac{
      V'(\phi)
    }{
      V(\phi)
    }
  \right)^2
  =
  f[V(\phi)]
  \mathcomma
\end{equation}
where we introduced the compact notation $V(\phi) \equiv \expval{\op{V}(\phi)}$ and $f[V(\phi)]$ is some functional to be specified later.
The method we will be employing to solve this problem is an algebraic approach introduced in \cite{BBC19} which we extend to noncommuting variables and connect to the Fock representation underlying the kinematics of \ac{gft}.

We work in the Heisenberg picture and assume that the Schrödinger and Heisenberg picture coincide at $\phi = 0$.
Operators without argument denote the Schrödinger picture operators.
The equations of motion for $\op{K}_0$ and $\op{K}_+ - \op{K}_-$ are given by ($\op{K}_+ + \op{K}_-$ is proportional to the Hamiltonian and therefore constant under time evolution)
\begin{eqnarray}
  &
  \label{eq:tm_k0_eom}
  \op{K}_0'
  (\phi)
  =
  \imagi \omega
  (
    \op{K}_+
    -
    \op{K}_-
  )
  (\phi)
  \mathcomma
  \\
  &
  (\op{K}_+ - \op{K}_-)'(\phi)
  =
  - 4 \imagi \omega \op{K}_0
\end{eqnarray}
which are solved by
\begin{eqnarray}
  \label{eq:tm_k0_heisenberg}
  &
  \op{K}_0(\phi)
  =
  \op{K}_0
  \cosh(2 \omega \phi)
  +
  \frac{\imagi}{2}
  (\op{K}_+ - \op{K}_-)
  \sinh(2 \omega \phi)
  \mathcomma
  \\
  &
  (\op{K}_+ - \op{K}_-)(\phi)
  =
  (\op{K}_+ - \op{K}_-)
  \cosh(2 \omega \phi)
  -
  2 \imagi \op{K}_0
  \sinh(2 \omega \phi)
  \mathperiod
\end{eqnarray}
From this one gets for the time dependence of the number operator
\begin{equation}
  \label{eq:tm_n_heisenberg}
  \op{N}(\phi)
  =
  -\frac{1}{2}
  +
  \left(
    \op{N}
    + \frac{1}{2} \op{I}
  \right)
  \cosh(2 \omega \phi)
  +
  \imagi
  (\op{K}_+ - \op{K}_-)
  \sinh(2 \omega \phi)
  \mathperiod
\end{equation}
The expectation value of this is nonnegative for all Fock states and grows exponentially at early or late times ($|\omega\phi|\gg 1$).
A nonvanishing expectation value $\expval{\imagi(\op{K}_+ - \op{K}_-)}$ implies a time asymmetry in the resulting effective cosmological history, i.e., different pre- and post-bounce phases.
For generic states $\expval{\op{N}(\phi)}$ is positive for all $\phi$; the only cases for which it becomes zero at some point during the evolution is for states which satisfy $ \abs{\expval{\op{K}_+ - \op{K}_-}}^2 \geq \expval{\op{N}} (\expval{\op{N}} + 1)$.
For Fock coherent states this is only the case for the Fock vacuum for which both sides are zero.
For both \ac{pg} and \ac{bg} coherent states it depends on the value of the Bargmann index $k$.
For $k > 1/4$ this inequality never holds, for $k < 1/4$, however, there are states for which the inequality is satisfied and for $k=1/4$ it is only the ground state (analogue to the Fock vacuum) for which the inequality holds.
\Eref{eq:tm_n_heisenberg} reproduces the result obtained in the `toy model' context of \cite{AGW18} (cf.~\eref{eq:gft_cosmo_toymodel_n} and below), the only difference being that a factor 5 is replaced by 1 since we consider a single field mode, not five modes as in the toy model.

One can derive an effective Friedmann equation directly by taking the expectation value of the explicit expression \eref{eq:tm_n_heisenberg}.
One then finds
\begin{equation}
  \label{eq:tm_eff_friedmann_from_heisenberg}
  \fl
  \left(
    \frac{
      V'(\phi)
    }{
      V(\phi)
    }
  \right)^2
  =
  4\omega^2
  \left(
    1
    +
    \frac{v_0}{V(\phi)}
    -
    \frac{v_0^2}{V(\phi)^2}
    \left[
      \expval{\op{N}}^2 + \expval{\op{N}}
      -
      \expval{\imagi(\op{K}_+ - \op{K}_-)}^2
    \right]
  \right)
  \mathcomma
\end{equation}
where $V(\phi) \equiv v_0 \expval{\op{N}(\phi)}$ (cf.~\eref{eq:tm_n_v_relation}).

One can, however, also obtain this effective Friedmann equation without knowing the time dependence of the number operator explicitly by using the algebraic structure of the system.
This was shown in \cite{BBC19} for the corresponding classical system, where the variables commute.
Here we extend this algebraic approach to the noncommutative case.

Starting from \eref{eq:tm_k0_eom} and the definition of the Casimir \eref{eq:su11_casimir} one arrives at
\begin{equation}
    \op{K}_0' (\phi)^2
  =
  4 \omega^2
  \left[
    \op{K}_0(\phi)^2
    -
    \left(
      \frac{\op{H}^2}{4 \omega^2}
      +
      \op{C}
    \right)
  \right]
\end{equation}
or written in terms of the number operator
\begin{equation}
  \label{eq:gft_alg_friedmann}
    \op{N}' (\phi)^2
  =
  4 \omega^2
  \left[
    \op{N}(\phi)^2
    +
    \op{N}(\phi)
    -
    \left(
      \frac{\op{H}^2}{\omega^2}
      +
      4 \op{C}
      -
      \frac{1}{4}
      \op{I}
    \right)
  \right]
\mathperiod
\end{equation}

In order to get the effective Friedmann equation one has to take the expectation value of \eref{eq:gft_alg_friedmann}.
However, it is crucial to note that in \eref{eq:tm_eff_fried_formal} the expectation value of the volume operator enters, rather than the expectation value of the volume operator squared.
The difference between the two is related to the variance of the volume, which is in general state-dependent.
Indeed, rearranging the expectation value of \eref{eq:gft_alg_friedmann} gives
\begin{equation}
  \label{eq:tm_eff_friedmann_k0}
  N'(\phi)^2
  =
  4 \omega^2
  \left[
    N(\phi)^2
    +
    N(\phi)
    +
    X
  \right]
\end{equation}
with $N(\phi) \equiv \expval{\op{N}(\phi)}$ and $X$ being given by
\begin{equation}
  X
  =
  \covariance{
    \op{N}
    (\phi)
  }{
    \op{N}
    (\phi)
  }
  -
  \frac{1}{4\omega^2}
  \covariance{
    \op{N}'
    (\phi)
  }{
    \op{N}'
    (\phi)
  }
  -
  \frac{
    \expval{\op{H}^2}
  }{
    \omega^2
  }
  -
  4 \expval{\op{C}}
  +
  \frac{1}{4}
  \mathcomma
\end{equation}
where the covariance $\covariance{\op{A}}{\op{B}}$ is defined as
\begin{equation}
  \covariance{
    \op{A}
  }{
    \op{B}
  }
  =
  \frac{1}{2}
  \expval{
    \op{A}
    \op{B}
    +
    \op{B}
    \op{A}
  }
  -
  \expval{\op{A}}
  \expval{\op{B}}
\end{equation}
(and for the case $\op{A}=\op{B}$ this would be called the variance of $\op{A}$).

We will now show that the quantity $X$ is indeed time-independent as suggested by the notation.
Noting that the variance of $\op{N}'(\phi)$ can be written as
\begin{equation}
  \covariance{
    \op{N}'
    (\phi)
  }{
    \op{N}'
    (\phi)
  }
  =
  - 4 \covariance{\op{H}}{\op{H}}
  + 16 \omega^2
  \covariance{
    \op{K}_+
    (\phi)
  }{
    \op{K}_-
    (\phi)
  }
  \mathcomma
\end{equation}
one can write $X$ as
\begin{equation}
  X
  =
  \covariance{
    \op{N}
    (\phi)
  }{
    \op{N}
    (\phi)
  }
  -
  4
  \covariance{
    \op{K}_+
    (\phi)
  }{
    \op{K}_-
    (\phi)
  }
  -
  \frac{
    \expval{\op{H}}^2
  }{
    \omega^2
  }
  -
  4 \expval{\op{C}}
  +
  \frac{1}{4}
  \mathperiod
\end{equation}
A quick calculation shows that
\begin{equation}
  \dfrac{\phi}
  \covariance{
    \op{K}_+
    (\phi)
  }{
    \op{K}_-
    (\phi)
  }
  =
  \frac{1}{2}
  \covariance{
    \op{N}'
    (\phi)
  }{
    \op{N}
    (\phi)
  }
\end{equation}
which shows that $X$ is time-independent, since both $\op{H}$ and $\op{C}$ are constants
of motion.
It follows that $X$ can be written as
\begin{equation}
  X
  =
  \covariance{
    \op{N}
  }{
    \op{N}
  }
  -
  4
  \covariance{
    \op{K}_+
  }{
    \op{K}_-
  }
  -
  \frac{
    \expval{\op{H}}^2
  }{
    \omega^2
  }
  -
  4
  \expval{\op{C}}
  +
  \frac{1}{4}
  \mathperiod
\end{equation}

Using the definition of the Casimir \eref{eq:su11_casimir} one can show that $X$ can be written in the alternative form
\begin{equation}
  X
  =
  - \expval{
    \op{N}
  }^2
  -
  \expval{
    \op{N}
  }
  +
  4
  \expval{
    \op{K}_+
  }
  \expval{
    \op{K}_-
  }
  -
  \frac{
    \expval{
      \op{H}
    }^2
  }{
    \omega^2
  }
  \mathperiod
\end{equation}
Yet another form can be obtained by inserting the Hamiltonian \eref{eq:tm_hamiltonian_su11} to get
\begin{equation}
  X
  =
  -
  \expval{
    \op{N}
    +
    \frac{1}{2}
    \op{I}
  }^2
  +
  \expval{
    \imagi
    (
      \op{K}_+
      -
      \op{K}_-
    )
  }^2
  +
  \frac{1}{4}
  \mathperiod
\end{equation}
The operators appearing in this last expression for $X$ are exactly those appearing in the formula for the operator $\op{K}_0(\phi)$ in the Heisenberg picture \eref{eq:tm_n_heisenberg} and one recovers the form of the effective Friedmann equation given in \eref{eq:tm_eff_friedmann_from_heisenberg}.
From the exact solution \eref{eq:tm_n_heisenberg}, one can see that $X\le 0$ in all Fock states, since $X>0$ would be equivalent to the number operator $N(\phi)$ taking a negative expectation value somewhere.
There are Fock states with $X=0$; these states encounter a singularity in their geometric interpretation, in the sense that the expectation value of the volume reaches zero somewhere and hence the effective energy density defined according to \eref{eq:tm_energy_density} diverges, even though the quantum evolution is completely regular even for these states.

Recalling that the volume operator is the rescaled number operator, i.e.\ $\op{V} = v_0 \op{N}$, one arrives at the effective Friedmann equation
\begin{equation}
  \label{eq:tm_eff_friedmann_full}
  \fl
  \left(
    \frac{
      V'(\phi)
    }{
      V(\phi)
    }
  \right)^2
  =
  4 \omega^2
  \left(
    1
    +
    \frac{
      v_0
    }{
      V(\phi)
    }
    -
    \frac{
      v_0^2
      \expval{
        \op{N}
      }
      (
        \expval{
          \op{N}
        }
        +
        1
      )
    }{
      V(\phi)^2
    }
    +
    \frac{
      4
      v_0^2
      \expval{
        \op{K}_+
      }
      \expval{
        \op{K}_-
      }
    }{
      V(\phi)^2
    }
    -
    \frac{
      v_0^2
      \expval{
        \op{H}
      }^2
    }{
      \omega^2
      V(\phi)^2
    }
  \right)
\end{equation}
Taking the late time limit (corresponding to large volumes) suggests that one should identify $12 \pi G := 4 \omega^2$ in order for the leading term to be compatible with the classical Friedmann dynamics.
This identification of fundamental couplings with an `emergent' Newton's constant is common in \ac{gft} cosmology \cite{OSW16,AGW18}.
Furthermore, identifying an energy density as in \eref{eq:tm_energy_density} and defining a critical energy density
\begin{equation}
  \rho_\critical
  =
  \frac{
    \omega^2
  }{
    2 v_0^2
  }
  =
  \frac{
    3 \pi G
  }{
    2 v_0^2
  }
  =
  \frac{3\pi}{2}
  \rho_\planck
  \left(
    \frac{v_\planck}{v_0}
  \right)^2
  \mathcomma
\end{equation}
where $\rho_\planck$ is the Planck mass density and $v_\planck$ is the Planck volume, the last term inside the parentheses in \eref{eq:tm_eff_friedmann_full} takes the form $-\rho/\rho_\critical$ familiar from \ac{lqc}.
The value for $\rho_\critical$ appearing here agrees with the critical density found in \cite{OSW16,OSW17}.
In close analogy to the results obtained in \ac{lqc} we then write for the effective Friedmann equation (see also \eref{eq:gft_cosmo_friedmann})
\begin{equation}
\label{eq:tm_friedman_with_rho_eff}
  \left(
    \frac{
      V'(\phi)
    }{
      V(\phi)
    }
  \right)^2
  =
  4 \omega^2
  \left(
    1
    -
    \frac{
      \rho_\effective(\phi)
    }{
      \rho_\critical
    }
  \right)
  +
  4 \omega^2
  \frac{
    v_0
  }{
    V(\phi)
  }
  \mathcomma
\end{equation}
where the effective energy density $\rho_\effective(\phi)$ is defined as
\begin{equation}
  \rho_\effective(\phi)
  =
  \rho_\phi(\phi)
  +
  \frac{
    \omega^2
    \expval{
      \op{N}
    }
    (
      \expval{
        \op{N}
      }
      +
      1
    )
  }{
    2V(\phi)^2
  }
  -
  \frac{
    2
    \omega^2
    \expval{
      \op{K}_+
    }
    \expval{
      \op{K}_-
    }
  }{
    V(\phi)^2
  }
  \mathperiod
\end{equation}
The first contribution to this effective energy density is given by the energy density $\rho_\phi(\phi)=\expval{\op{H}}^2/(2V(\phi)^2)$ associated to a massless scalar field (as defined in  \eref{eq:tm_energy_density}), but there are two additional contributions depending on the expectation values $\expval{\op{N}},\,\expval{\op{K}_+}$ and
    $\expval{
      \op{K}_-
    }$
in the initial state.
As these additional contributions to $\rho_\effective$ also scale as $V(\phi)^{-2}$, their effect is equivalent to a shift in the scalar field momentum compared to its classical value $\expval{\op{H}}$.
The last term in \eref{eq:tm_friedman_with_rho_eff}, scaling as $1/V(\phi)$, is similar to a correction found in mean-field calculations and takes the form of an effective matter contribution for matter with equation of state $p=2\rho$ (cf.~\cite{CPS16}).

Depending on the initial state, the quantity $X$ (or, alternatively, the additional contributions to the effective energy density) can take different forms.
For the \ac{pg} coherent states one finds the following form
\begin{equation}
  \label{eq:gft_su11_x_pg}
  X_{\mathsubscript{PG}}
  =
  - 4 k^2
  \frac{
    (1 + \zeta^2)
    (1 + \compconj{\zeta}^2)
  }{
    (1 - \abs{\zeta}^2)^2
  }
  +
  \frac{1}{4}
  =
  -
  4 k^2
  -
  \frac{
    \expval[\mathsubscript{PG}]{
      \op{H}
    }^2
  }{
    \omega^2
  }
  +
  \frac{1}{4}
  \mathperiod
\end{equation}
We note that for $k = 1/4$ this reduces to
$
  X_{\mathsubscript{PG}}
  =
  -
  \expval[\mathsubscript{PG}]{
    \op{H}
  }^2
  /
  \omega^2
$, and the effective energy density and classical energy density exactly coincide.
The effective Friedmann equation for \ac{pg} coherent states therefore reads
(for general $k$)
\begin{equation}
  \left(
    \frac{
      V'(\phi)
    }{
      V(\phi)
    }
  \right)^2
  =
  4 \omega^2
  \left(
    1
    +
    \frac{
      v_0
    }{
      V(\phi)
    }
    -
    \frac{
      v_0^2
      \expval[\mathsubscript{PG}]{
        \op{H}
      }^2
    }{
      \omega^2
      V(\phi)^2
    }
    - \frac{
      v_0^2\left(16 k^2
    - 1\right)
    }{
     4V(\phi)^2
    }
  \right)
  \mathperiod
\end{equation}

For Fock coherent states one gets for $X$
\begin{equation}
  X_{\mathsubscript{F}}
  =
  - \frac{1}{4} (\sigma^2 + \compconj{\sigma}^2)^2
  - \abs{\sigma}^2
  =
  -
  \expval[\mathsubscript{F}]{
    \op{N}
  }
  -
  \frac{
    \expval[\mathsubscript{F}]{
      \op{H}
    }^2
  }{
    \omega^2
  }
  \mathperiod
\end{equation}
Therefore the Friedmann equation for Fock coherent states is given by
\begin{equation}
  \label{eq:tm_eff_fried_fock}
  \left(
    \frac{
      V'(\phi)
    }{
      V(\phi)
    }
  \right)^2
  =
  4 \omega^2
  \left(
    1
    +
    \frac{
      v_0
    }{
      V(\phi)
    }
    -
    \frac{
      v_0^2
      \expval[\mathsubscript{F}]{
        \op{N}
      }
    }{
      V(\phi)^2
    }
    -
    \frac{
      v_0^2
      \expval[\mathsubscript{F}]{
        \op{H}
      }^2
    }{
      \omega^2
      V(\phi)^2
    }
  \right)
  \mathperiod
\end{equation}

For completeness we also state the value of $X$ one gets for \ac{bg} coherent states,
\begin{equation}
  X_{\mathsubscript{BG}}
  =
  -
  (\chi - \compconj{\chi})^2
  -
  4
  \left(
    k
    +
    \abs{\chi}
    \frac{
      I_{2k}(2 \abs{\chi})
    }{
      I_{2k - 1}(2 \abs{\chi})
    }
  \right)^2
  +
  \frac{1}{4}
  \mathperiod
\end{equation}

The Friedmann equations derived here are compatible with previous results in \ac{gft} cosmology \cite{OSW16,OSW17,AGW18,WEw19} where either a mean-field approach was used or a simplifying assumption was imposed on the initial conditions.
We emphasise that no approximations were used which resulted in the appearance of extra terms.
In particular, we were able to identify one of those extra terms with the energy density of the real scalar field acting as a relational clock variable.
Corrections to the classical Friedmann dynamics of the \ac{lqc}-like form $-\rho/\rho_\critical$, which lead to effective repulsive behaviour and a bounce at high energies, were found for all coherent states considered.
We also found that in general the `effective' energy density appearing in the Friedmann equation \eref{eq:tm_friedman_with_rho_eff} is not equal to the classical energy density $\rho_\phi=\pi_\phi^2/(2V^2)$ of a massless scalar field, but contains additional terms depending on the initial conditions chosen.
 
\section{Interacting toy model}
In this section we extend the toy model discussed in \sref{sec:tm} by adding an interaction term to the Hamiltonian.
The resulting interacting model still represents a simplification of the dynamics of full \ac{gft}, since we continue to assume that only one mode is relevant.
While the general expectation is that the dynamics should depend on the coupling of different modes, studying this simpler model can provide insights on how \ac{gft} interactions can change the interpretation of the dynamics in terms of effective cosmology.
A similar model, which included polynomial interactions for a single \ac{gft} field mode, was previously studied in \cite{CPS16} in a mean-field approximation (see also \cite{PS17}), leading to corrections to the effective Friedmann equations coming from these interactions.
These corrections become more significant at late times as the impact of \ac{gft} interactions grows with the number of quanta.
Here we will be able to contrast these mean-field results with effective modified Friedmann equations obtained in a more general setting.

We now consider a Hamiltonian given in terms of the $\liealg{su}(1, 1)$ variables by
\begin{equation}
  \label{eq:int_hamiltonian}
  \op{H}
  =
  -\omega
  (
    \op{K}_+
    + \op{K}_-
  )
  +
  \lambda \omega
  (
    \op{K}_+
    + \op{K}_-
    + 2 \op{K}_0
  )^2
  \mathperiod
\end{equation}
In terms of the bosonic realisation of the $\liealg{su}(1, 1)$ algebra \eref{eq:su11_bosonic_realisation} the Hamiltonian reads
\begin{equation}
  \label{eq:int_hamiltonian_fock_vars}
  \op{H}
  =
  - \frac{\omega}{2}
  (
    \op{a}^2
    + \hermconj{\op{a}}{}^2
  )
  +
  \frac{
    \lambda \omega
  }{
    4
  }
  (
    \op{a}
    + \hermconj{\op{a}}
  )^4
  \mathperiod
\end{equation}
Recalling the definitions of the creation and annihilation operators in terms of the \ac{gft} field and its conjugate momentum \eref{eq:gft_cosmo_annihilation}, \eref{eq:gft_cosmo_creation} one can rewrite the Hamiltonian as
\begin{equation}
  \label{eq:int_hamiltonian_gft_vars}
  \op{H}
  =
  \frac{1}{2 \abs{\mathcal{K}^{(2)}}}
  \op{\pi}^2
  -
  \frac{1}{2}
  \abs{\mathcal{K}^{(0)}}
  \op{\varphi}^2
  +
  \lambda
  \abs{\mathcal{K}^{(0)}}^{3/2}
  \abs{\mathcal{K}^{(2)}}^{1/2}
  \op{\varphi}^4
  \mathcomma
\end{equation}
where we suppressed the Peter--Weyl representation labels and we also assume the mode to be of the type discussed at the end of \sref{sec:gft_cosmo} with magnetic indices $m_i = 0$.
From this expression one sees that the interaction term would correspond to a $\varphi^4$ interaction term in an appropriately defined \ac{gft} action.

The dynamics of this system crucially depend on the sign of $\lambda$.
Indeed, for positive $\lambda$ this Hamiltonian will be bounded from below, whereas for the case that $\lambda$ is negative the Hamiltonian is unbounded as it is in the free case.
Interpreting \eref{eq:int_hamiltonian_gft_vars} as a mechanical system with kinetic and potential terms, one sees that for positive $\lambda$ one gets a `Mexican hat' type potential, whereas in the case of negative $\lambda$ the potential is an `upside-down' anharmonic oscillator.
In the cosmological context one expects from this that for negative $\lambda$ the Universe will undergo an enhanced exponential expansion and for positive $\lambda$ the Universe will recollapse after some time leading to a cyclic cosmology.
Such a cyclic cosmology was indeed found in \cite{CPS16}, where $\lambda > 0$ was assumed.
In \sref{sec:int_alg_poisson} we argue that this expectation is correct if \eref{eq:int_hamiltonian} is seen as the Hamiltonian of a classical system.

When one tries to use the algebraic approach detailed in \sref{sec:tm} to derive an effective Friedmann equation for the interacting model \eref{eq:int_hamiltonian} one faces several challenges.
Firstly, the noncommutativity does not allow the reduction to a small set of `basis operators'.
Secondly, the expressions involved feature products of three operators and it is technically challenging to relate them to the expectation values of `simple' operators such as the Hamiltonian and number operator.

To begin with, we restrict ourselves to the classical case, where the variables commute and one can employ the algebraic approach to derive an effective Friedmann equation.
We find an exact Friedmann equation whose limits at early and late times are given.
After that we turn to the general (quantum) case, where the operators do not commute.
In that case we resort to a perturbative treatment which is valid at early times.
Furthermore, we perform a numerical analysis for Fock coherent states.
We find that a linear (perturbative) correction to the effective Friedmann equation can capture the effect of the interaction term for a short time, after which the dynamics become nonperturbative.

\subsection{Algebraic approach for classical analogue system}
\label{sec:int_alg_poisson}
In this section we study a classical dynamical system with time evolution generated by the Hamiltonian \eref{eq:int_hamiltonian}.
In this approach the $\liealg{su}(1,1)$ variables $K_0$, $K_+$ and $K_-$ are not viewed as quantum operators but as coordinates on a Poisson manifold, subject to an $\liealg{su}(1,1)$ Poisson algebra which then defines the Hamiltonian dynamics.
In contrast to the full quantum case, these variables commute and we interpret the variables themselves as the observables of interest, i.e.\ one does not have to take expectation values.
In this sense, this approximation neglects all quantum corrections coming from operator orderings and uncertainties in a quantum state.
We will identify the variable $K_0$ with the particle number $N$ such that $N \equiv 2 K_0$.
As above, we assume the total volume to be proportional to the particle number, $V = v_0 N$, and switch between $N$ and $V$ freely.

The equation of motion of the variable $N$ is given by
\begin{equation}
  N'(\phi)
  =
  2 \imagi \omega
  (K_+(\phi) - K_-(\phi))
  (
    1 - 2 \lambda (K_+(\phi) + K_-(\phi) + N(\phi))
  )
  \mathperiod
\end{equation}
After squaring this equation one can use the Casimir \eref{eq:su11_casimir} to replace the combination $(K_+ - K_-)$.
One is then left with an expression where only the combination $(K_+ + K_-)$ appears,
\begin{equation}
  \label{eq:int_classical_eff_friedmann_intermediate}
  \eqalign{
    N'(\phi)^2
    =
    &
    4 \omega^2
    \left(
      N(\phi)^2
      - 4C
      - (K_+(\phi) + K_-(\phi))^2
    \right)
    \\
    &
    \qquad
    \times
    \left(
      1
      - 2 \lambda (K_+(\phi) + K_-(\phi) + N(\phi))
    \right)^2
    \mathperiod
  }
\end{equation}
From the Hamiltonian one can derive an explicit formula for $(K_+ + K_-)$
\begin{equation}
  \label{eq:int_kP_plus_kM}
  K_+(\phi)
  +
  K_-(\phi)
  =
  \frac{1}{2 \lambda}
  \left(
    1
    - 2 \lambda N(\phi)
    -
    \sqrt{
      1
      + 4 \lambda \left( \frac{H}{\omega} - N(\phi) \right)
    }
  \right)
  \mathcomma
\end{equation}
where we chose the solution connected to the free theory.
Inserting this into \eref{eq:int_classical_eff_friedmann_intermediate} one gets the nonperturbative effective Friedmann equation
\begin{equation}
  \label{eq:int_classical_eff_friedmann_nonpert}
  \fl
\eqalign{
  N'(\phi)^2
  =
  &
  -
  \frac{
    2
    \omega^2
  }{
    \lambda^2
  }
  \left(
    1 + 4 \lambda
    \left(
      \frac{H}{\omega} - N(\phi)
    \right)
  \right)
    \times
    \Bigg[
      1 - 4 \lambda N(\phi)
  \\
  &
  \qquad
  \qquad
      - (1 - 2 \lambda N(\phi))
      \sqrt{
        1 + 4 \lambda
        \left(
          \frac{H}{\omega} - N(\phi)
        \right)
      }
      + 2 \lambda
      \left(
        4 \lambda C
        + \frac{H}{\omega}
      \right)
    \Bigg]
    \mathperiod
}
\end{equation}
We already see that $\lambda > 0$ implies an upper bound on the value of $N$ and hence the volume, since the right-hand side of \eref{eq:int_classical_eff_friedmann_nonpert} has to be real and positive.
Indeed, at exactly that upper limit the right-hand side of \eref{eq:int_classical_eff_friedmann_nonpert} becomes zero leading to a recollapse.
In the case $\lambda < 0$, for the right-hand side of \eref{eq:int_classical_eff_friedmann_nonpert} to be real and positive $N$ has to be greater than some minimal value (which can be zero).
The right-hand side remains real and positive for all values of $N$ greater than that minimal value, implying that the Universe expands indefinitely.
For clarity, we recall that $H$ here and throughout the paper denotes the Hamiltonian or energy (interpreted as the canonical momentum conjugate to the scalar field $\phi$), not a Hubble rate in cosmology.

We would like to interpret \eref{eq:int_classical_eff_friedmann_nonpert} in terms of a cosmological model given by one or several matter components which contribute to the matter energy density on the right-hand side.
For a general such model in usual classical cosmology, with $n$ matter components labelled by an index $i$ viewed as perfect fluids with each having an equation of state $p_i=w_i \rho_i$, the Friedmann equation for the volume as a function of relational time $\phi$ would be of the form
\begin{equation}
  \label{eq:int_friedmann_eos}
  \left(
    \frac{V'(\phi)}{V(\phi)}
  \right)^2
  = \sum_{i=1}^n
  A_i V(\phi)^{1-w_i}
\end{equation}
where $A_i$ are constants of motion.
While \eref{eq:int_classical_eff_friedmann_nonpert} is valid at all times, its interpretation in terms of cosmological models of the form \eref{eq:int_friedmann_eos} is not clear due to the appearance of a square root on the right-hand side.
Moreover, additional matter components as in \eref{eq:int_friedmann_eos} would come with new conserved quantities $A_i$, whose values can be varied independently (or, in other words, are determined by the initial conditions).
In our model, the presence of \ac{gft} interactions does not introduce new parameters set by initial conditions, only a new coupling constant $\lambda$; these interactions therefore modify the dynamics of gravity rather than matter.
Here we follow the convention in which quantum gravity corrections are written as modifying the right-hand side rather than the left-hand side of Friedmann equations (as is usually done in \ac{lqc}) in order to give intuition for the effective dynamics.\footnote{An alternative possible interpretation of effective Friedmann equations obtained from quantum gravity models is to view them as equivalent to Friedmann equations of a modified theory of gravity.
  A general reconstruction method of this type for mimetic gravity theories was developed in \cite{Ces19}.
}

One might be interested in the Friedmann equation valid at relatively small volumes where interactions have become relevant but not dominant, so that one can employ perturbation theory.
Expanding \eref{eq:int_classical_eff_friedmann_nonpert} as a series around $\lambda=0$ one gets
\begin{equation}
  \label{eq:int_classical_eff_friedmann_series}
  \fl
  \eqalign{
    N'(\phi)^2
    =
    4\omega^2
    \Bigg\{
  &
      N(\phi)^2
      - \frac{H^2}{\omega^2}
      - 4C
      \left(
        1
        -
        4 \lambda
        \left(
          N(\phi) - \frac{H}{\omega}
        \right)
      \right)
  \\
  &
    \eqalign{
      -
      \sum_{n=1}^\infty
      \bigg[
    &
      12 \lambda^n
      \frac{
        (2n - 2)!
      }{
        (n - 1)!
        (n + 2)!
      }
      \left(
        (2n - 1) \frac{H}{\omega}
        + (3 - n) N(\phi)
      \right)
    \\
    &
      \times
      \left(
        N(\phi)
        - \frac{H}{\omega}
      \right)^{n + 1}
      \bigg]
      \Bigg\}
      \mathperiod
    }
  }
\end{equation}
From this form one can see that it is the product $\lambda N(\phi)$ that must be small for the perturbative expansion to make sense.
Comparing with \eref{eq:int_friedmann_eos}, one could interpret the leading (linear) correction coming from the interaction term as an effective matter component with an equation of state parameter $w=0$, i.e., a dust component.
This result would then agree with the results of a mean-field calculation in \cite{CPS16} where adding a $\varphi^4$ interaction to the \ac{gft} Lagrangian led to such a dust-like contribution in the effective cosmology.
However, the full expansion given in \eref{eq:int_classical_eff_friedmann_series} shows that such an interpretation would only be valid in an intermediate regime in which the product $\lambda N(\phi)$ is no longer negligible, but also not yet large enough for higher orders to contribute.
Indeed, as the volume grows further, $\lambda N(\phi)$ would soon be $\bigO{1}$ and the perturbative expansion receives contributions from \emph{all} orders.
In particular, there is never a regime in which the effective Friedmann equation is dominated by the `dust-like' component, as it would be in the mean-field form obtained in \cite{CPS16}.
One possible interpretation of this discrepancy is that mean-field methods are strictly only valid in the free theory, since they assume the absence of correlations between quanta.
Hence one would expect them to become inaccurate as the contribution to the effective dynamics coming from \ac{gft} interactions becomes strong.
Care has to be taken when interpreting \eref{eq:int_classical_eff_friedmann_series} up to some order.
For instance, if one only considers terms up to order $\lambda^2$ in \eref{eq:int_classical_eff_friedmann_series}, one would conclude that there is always a recollapse (even for negative $\lambda$), which is not true for the full solution.

To understand the late-time behaviour of the system, recall that $N(\phi)$ grows without bound while the other dynamical quantities on the right-hand side of \eref{eq:int_classical_eff_friedmann_nonpert}, $C$ and $H$, are constants of motion.
In the limit of large $N(\phi)$ (corresponding to late times) the leading order contribution of \eref{eq:int_classical_eff_friedmann_nonpert} is given by
\begin{equation}
  \label{eq:int_classical_eff_friedmann_asymptotic}
  \left(
    \frac{N'(\phi)}{N(\phi)}
  \right)^2
  =
  32 \omega^2
  \sqrt{-\lambda N(\phi)}
  +
  \bigO{1}
  \mathperiod
\end{equation}
Interpreting the right-hand side of \eref{eq:int_classical_eff_friedmann_asymptotic} as an energy density of matter, one finds from \eref{eq:int_friedmann_eos} that it corresponds to matter with an equation of state parameter $w = 1/2$.
The solutions to this asymptotic form of the effective Friedmann equation behave as $N(\phi)\sim|\phi-\phi_0|^{-4}$, diverging at some $\phi=\phi_0$.
We would hence expect the evolution of our system to terminate at some finite value of $\phi$, depending on initial conditions.
Notice that in this interpretation the energy density of this new `matter' is fixed by the \ac{gft} coupling $\lambda$ and hence, as we mentioned before, the effect of adding \ac{gft} interactions should rather be seen as modifying the dynamics of gravity on large scales.
The scale at which such a modification would become relevant depends on the chosen value of $\lambda$.

\begin{figure}[htpb]
  \centering
  \includegraphics{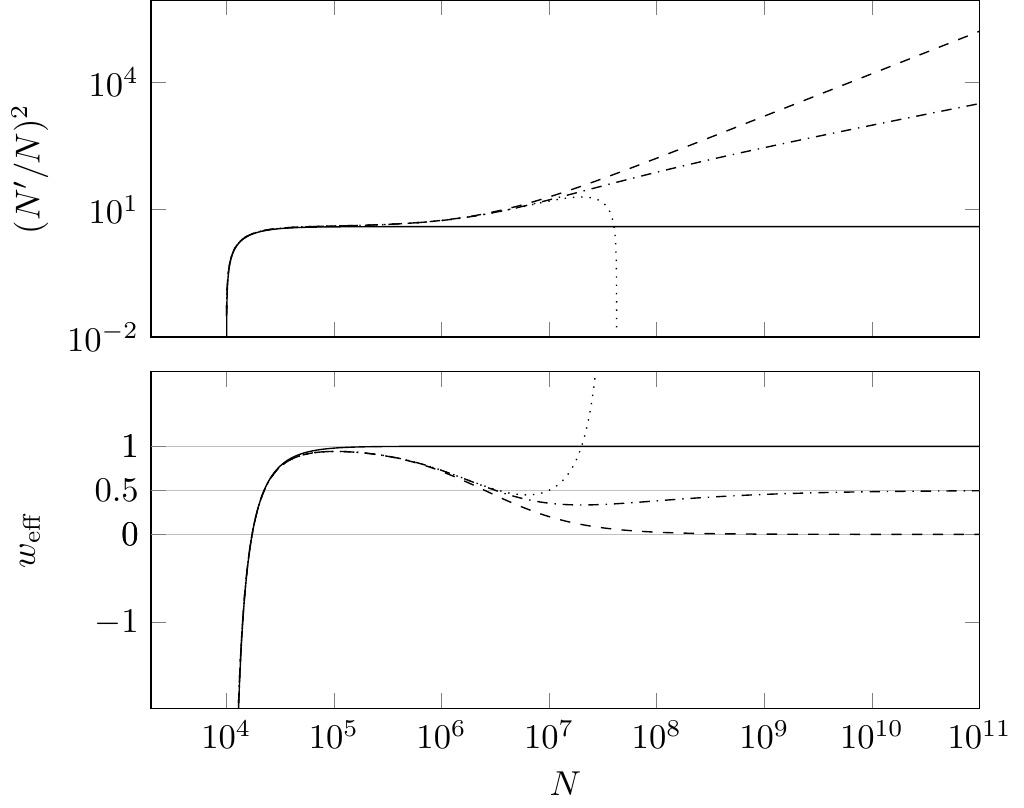}
  \caption{The relative relational expansion rate squared \eref{eq:int_classical_eff_friedmann_nonpert} and the `effective equation of state parameter' \eref{eq:int_eos_eff} as functions of $N$.
    The solid lines correspond to a truncation at zeroth order in $\lambda$.
    The dashed lines correspond to a truncation at first order in $\lambda$.
    The dotted lines correspond to a truncation at second order in $\lambda$.
    The dash-dotted lines correspond to the full nonperturbative case.
    The parameters are:
    $\omega = 1$,
    $\lambda = - 10^{-7}$,
    $H = -10040$,
    $C = - 3/16$.
    (The choice of $H$ corresponds to $\bra{\sigma} \op{H} \ket{\sigma}$ for $\sigma = 100$.)
  }
  \label{fig:int_eos_poisson}
\end{figure}

To characterise the dynamics of the effective cosmology at arbitrary times, from \eref{eq:int_friedmann_eos} we define an `effective equation of state parameter'
\begin{equation}
  \label{eq:int_eos_eff}
  w_{\effective}(\phi)
  =
  1
  -
  \frac{
    \dd \log \left( N'(\phi)/N(\phi) \right)^2
  }{
    \dd \log N(\phi)
  }
\end{equation}
which is plotted in \fref{fig:int_eos_poisson} as a function of $N(\phi)$ for various truncations of \eref{eq:int_classical_eff_friedmann_series}.
In the plot one sees that for $\abs{\lambda N(\phi)} \lesssim 10^{-2}$ the interactions become relevant and that for $\abs{\lambda N(\phi)} \lesssim 1$ the difference between the exact result and the first order truncation becomes large as expected.
Another observation is that while for small values of $\abs{\lambda N(\phi)}$ the second order truncation is in better agreement with the nonperturbative results, this truncation quickly diverges for $\abs{\lambda N(\phi)} \gtrsim 1$.

\subsection{Quantum calculation}
We are not able to derive an exact solution for $\hat{N}(\phi)$ in the interacting quantum mechanical case, where operators do not commute and higher moments and simple expectation values are independent.
To give approximate solutions, we first give a perturbative analytical method which we then contrast with numerical results.
We expand the number operator as a series with expansion parameter $\lambda$, i.e.\
\begin{equation}
  \label{eq:int_n_pert_series}
  \op{N}(\phi)
  =
  \sum_{n=0}^\infty
  \lambda^n
  \op{N}_n(\phi)
  \mathperiod
\end{equation}
Splitting the Hamiltonian $\op{H}$ into a $\lambda$-independent part, $\op{H}_0$, and a $\lambda$-dependent part, $\op{H}_1$, one can split the time evolution operator in a way similar to what is done in the interaction picture as $\op{U}(\phi) = \op{U}_0(\phi) \op{U}_\interaction(\phi)$, where $\op{U}_0(\phi) = \exp(-\imagi \op{H}_0 \phi)$ is the time evolution operator of the free system and the interaction time evolution operator is defined by the time-ordered exponential
\begin{equation}
  \op{U}_\interaction(\phi)
  =
  \timeordering
  \exp
  \left(
    -\imagi
    \int_0^\phi
    \intmeasure{\phi'}\,
    \op{U}_0^{-1}(\phi')
    \op{H}_1
    \op{U}_0(\phi')
  \right)
  \mathperiod
\end{equation}
Note that the inverse interaction time evolution operator requires anti-time-ordering.
Expanding the interaction time evolution operator as a series in $\lambda$,
\begin{equation}
  \op{U}_\interaction(\phi)
  =
  \sum_{n=0}^\infty
  \lambda^n
  (\op{U}_\interaction)_n(\phi)
  \mathcomma
\end{equation}
and inserting into \eref{eq:int_n_pert_series} gives for the $\op{N}_n(\phi)$
\begin{equation}
  \op{N}_n(\phi)
  =
  \sum_{m=0}^n
  (\op{U}_\interaction^{-1})_m(\phi)
  \op{U}_0^{-1}(\phi)
  \op{N}
  \op{U}_0(\phi)
  (\op{U}_\interaction)_{n-m}(\phi)
  \mathperiod
\end{equation}
The strategy would be to find the exact form of these $\op{N}_n(\phi)$ up to some order and then derive an effective Friedmann equation from these.
We will only do this to first order.

The leading term $\op{N}_0(\phi)$ is of course the same as in the free theory \eref{eq:tm_n_heisenberg}.
The first order term $\op{N}_1(\phi)$ is given by
\begin{equation}
\eqalign{
  \fl
  \op{N}_1(\phi)
  =
  \frac{\lambda}{2}
  \Big\{
    &
    \mkern-12mu (\op{K}_+)^2
    \left[
      3
      - 2 (2 + 3\imagi\omega\phi) \cosh(2\omega\phi)
      + \cosh(4\omega\phi)
    \right]
    + \hermconjtext
    \\
    &
    +
    \op{K}_+
    (2\op{N} + 3)
    \left[
      (\imagi - 3\omega\phi) \sinh(2\omega\phi)
      - \imagi \sinh(4\omega\phi)
    \right]
    + \hermconjtext
    \\
    &
    -
    \frac{1}{2}
    \sinh^2(2\omega\phi)
    [
      3 + 4 \op{N}^2 + 8 \op{N} + 8 \op{K}_+ \op{K}_-
    ]
  \Big\}
  \mathperiod
}
\end{equation}

As outlined above, what one would like to do is take the expectation values of the perturbative expansion \eref{eq:int_n_pert_series} and derive an effective Friedmann equation for arbitrary states, valid up to some order in $\lambda$.
However, already the first order expression for $\op{N}(\phi)$ is quite complicated and we were not able to derive a corresponding effective Friedmann equation for general states.
To ameliorate this we resort to taking the expectation values for some specific classes of coherent states.

Firstly, we turn to the Fock coherent states.
Writing for the parameter of the Fock coherent states \eref{eq:tm_coh_state_fock} $\sigma = \sigma_1 + \imagi \sigma_2$, one finds for the case $\sigma_2 = 0$, i.e.\ for real $\sigma$,
\begin{equation}
  \label{eq:int_eff_fried_fock_real_order1}
  \fl
  \eqalign{
    N'(\phi)^2
    =
    4 \omega^2
    \bigg(
    &
    N(\phi)^2
    + N(\phi)
    - \sigma_1^2 (1 + \sigma_1^2)
    \\
    &
    \eqalign{
      +
      \frac{
        \lambda
      }{
        (1 + 2 \sigma_1^2)^2
      }
      &
      \Big[
        -4 N(\phi)^3 (3 + 12 \sigma_1^2 + 4 \sigma_1^4)
        \\
        &
        - 6 N(\phi)^2 (3 + 15 \sigma_1^2 + 12 \sigma_1^4 + 4 \sigma_1^6)
        \\
        &
        + 6 N(\phi) (-1 - 5 \sigma_1^2 + 4 \sigma_1^6)
        \\
        &
        +  2 \sigma_1^2 (3 + 24 \sigma_1^2 + 51 \sigma_1^4 + 48 \sigma_1^6 +
        20 \sigma_1^8)
      \Big]
      + \bigO{\lambda^2}
      \bigg)
      \mathperiod
    }
  }
\end{equation}
For the case $\sigma_1 = 0$, i.e.\ imaginary $\sigma$, one finds the similar expression
\begin{equation}
  \fl
  \eqalign{
    N'(\phi)^2
    =
    4 \omega^2
    \bigg(
    &
    N(\phi)^2
    + N(\phi)
    - \sigma_2^2 (1 + \sigma_2^2)
    \\
    &
    \eqalign{
      +
      \frac{
        \lambda
      }{
        (1 + 2 \sigma_2^2)^2
      }
      &
      \Big[
        -4 N(\phi)^3 (3 + 12 \sigma_2^2 + 4 \sigma_2^4)
        \\
        &
        - 6 N(\phi)^2 (3 + 9 \sigma_2^2 - 4 \sigma_2^4 - 4 \sigma_2^6)
        \\
        &
        + 6 N(\phi) (-1 + \sigma_2^2 + 16 \sigma_2^4 + 12 \sigma_2^6)
        \\
        &
        +  2 \sigma_2^2 (3 + 6 \sigma_2^2 - 15 \sigma_2^4 - 24 \sigma_2^6
          - 4 \sigma_2^8)
      \Big]
      + \bigO{\lambda^2}
      \bigg)
      \mathperiod
    }
  }
\end{equation}
The expression for the general case, where $\sigma$ can be any complex number, is quite involved and we do not state it here.

Secondly, we turn to the \ac{pg} coherent states.
For the these states one finds the following general expression
\begin{equation}
  \fl
  \eqalign{
  N'(\phi)^2
  =
  4 \omega^2
  \Bigg\{
  &
    \left(
      N(\phi)
      + \frac{1}{2}
    \right)^2
    - 4 k^2
    - \frac{\expval[\mathsubscript{PG}]{\op{H}}^2}{\omega^2}
  \\
  &
    \eqalign{
    +
    \lambda
    \frac{2k + 1}{4k}
    \Big[
    &
      - 8 N(\phi)^3
      + 12 N(\phi)^2
      \left(
        \frac{\expval[\mathsubscript{PG}]{\op{H}}}{\omega} - 1
      \right)
    \\
    &
      + 2 N(\phi)
      \left(
        6 \frac{\expval[\mathsubscript{PG}]{\op{H}}}{\omega}
        + 16 k^2
        - 3
      \right)
    \\
    &
      -
      \left(
        \frac{\expval[\mathsubscript{PG}]{\op{H}}}{\omega}
        - 1
      \right)
      \left(
        2 \frac{\expval[\mathsubscript{PG}]{\op{H}}^2}{\omega^2}
        + \frac{\expval[\mathsubscript{PG}]{\op{H}}}{\omega}
        + 16 k^2
        - 1
      \right)
    \Big]
    }
  \\
  &
    +
    \bigO{\lambda^2}
    \Bigg\}
    \mathperiod
  }
\end{equation}
It is remarkable that it is possible to write the right-hand side only in terms of $N(\phi)$ and $\expval[\mathsubscript{PG}]{H}$.
As before, we would expect all higher order corrections to become relevant as soon as a regime is reached in which $\abs{\lambda N(\phi)} \lesssim 1$.

\begin{figure}[htpb]
  \centering
  \includegraphics{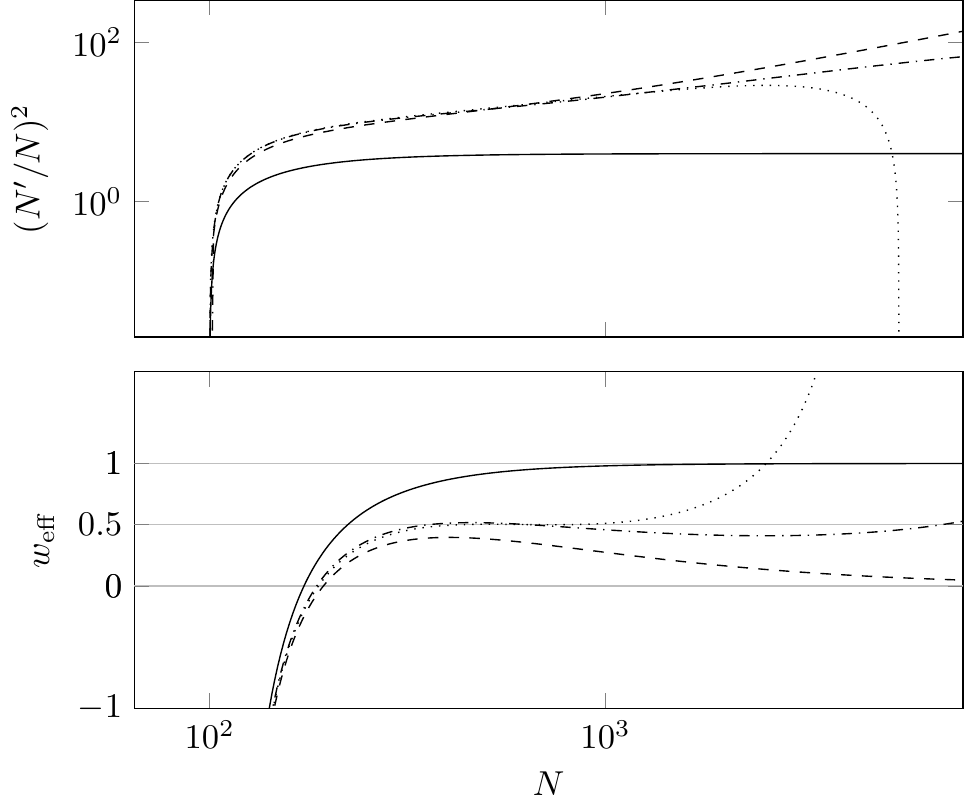}
  \caption{
    The relative relational expansion rate squared and the `effective equation of state parameter' \eref{eq:int_eos_eff} as functions of $N$ for Fock coherent states with real parameter $\sigma$.
    The solid lines correspond to a truncation at zeroth order in $\lambda$.
    The dashed lines correspond to a truncation at first order in $\lambda$.
    The dotted lines correspond to a truncation at second order in $\lambda$.
    The dash-dotted lines correspond to the full nonperturbative case.
    The parameters are:
    $\omega = 1$,
    $\lambda = - 10^{-3}$,
    $\sigma = 10$
  }\label{fig:int_quantum}
\end{figure}

In \fref{fig:int_quantum} the relative relational expansion rate squared and the effective equation of state parameter \eref{eq:int_eos_eff} are plotted as a function of $N$ for different truncations of the perturbative expansion and for the result of a numerical calculation.\footnote{The numerical results were obtained by solving the time-dependent Schrödinger equation of the Hamiltonian \eref{eq:int_hamiltonian_fock_vars} in the position representation.}
The zeroth order truncation corresponds to the free case and corresponds to \eref{eq:tm_eff_fried_fock} and the first order truncation is given in \eref{eq:int_eff_fried_fock_real_order1}.
Note that we do not state the second order truncation explicitly, since the expression is rather convoluted.
The state considered in this plot is a Fock coherent state with real parameter $\sigma$.
The numerical results are in good agreement with the second order truncation.
However, the second order truncation diverges quickly for values $\abs{\lambda N(\phi)} \gtrsim 1$, whereas the first order truncation does not.
Note that the parameters chosen do not accommodate a regime in which $\abs{\lambda N(\phi)} \ll 1$, explaining the mismatch with the linearised theory.
Due to numerical limitations we were not able to enter the asymptotic regime and determine the corresponding effective equation of state parameter for late times.

\begin{figure}[htbp]
  \centering
  \includegraphics{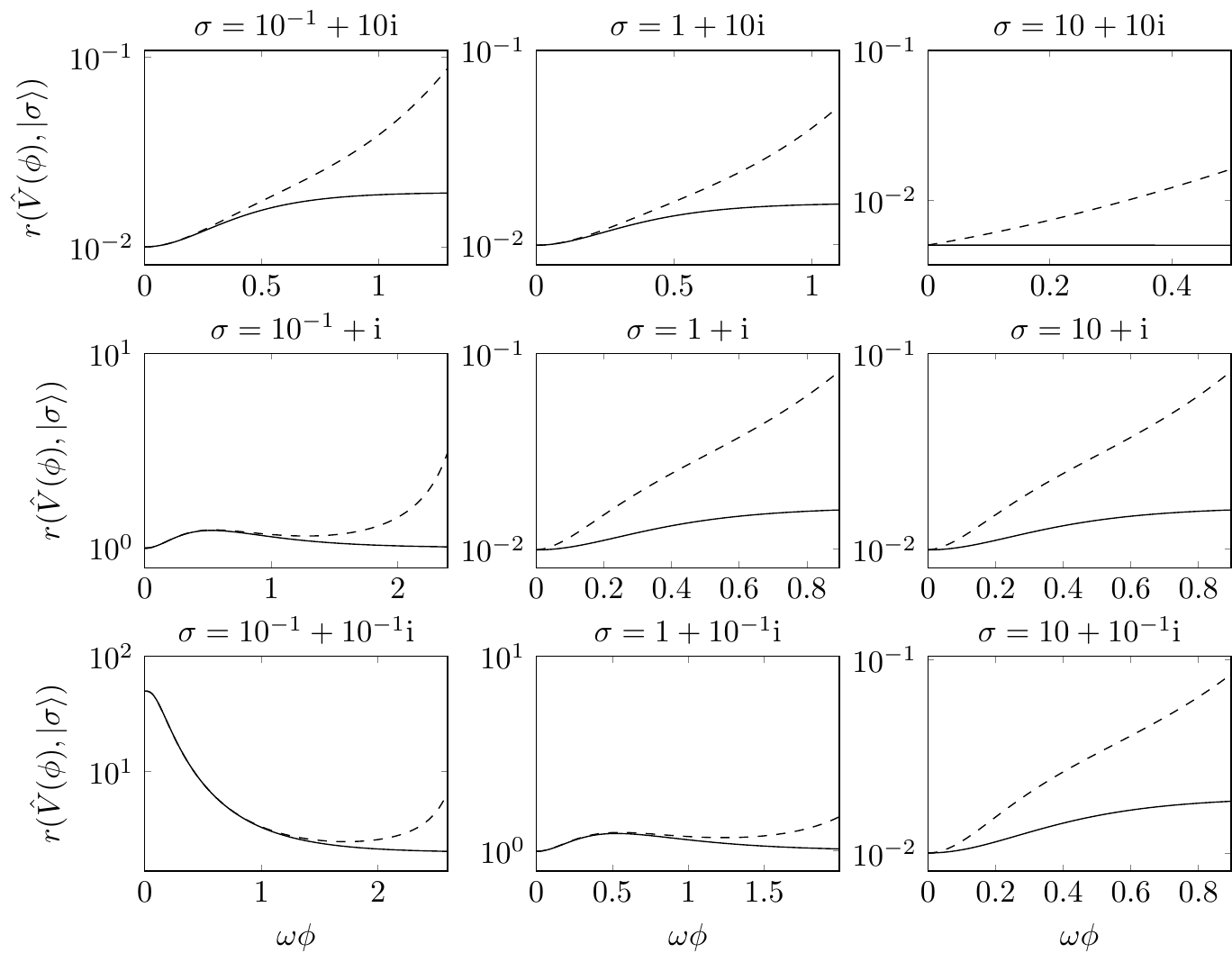}
  \caption{The relative uncertainty of the volume operator as a function of $\omega \phi$ for Fock coherent states.
    The solid line corresponds to the free case, $\lambda = 0$.
    The dashed line corresponds to the interacting case with $\lambda = -10^{-3}$.
  }
  \label{fig:int_rel_uncertainty_comp}
\end{figure}

We conclude the present discussion of the quantum behaviour of the interacting toy model by revisiting the relative uncertainties of the volume as a function of relational time presented in \sref{sec:tm_class_of_coh_stat_and_rel_uncert} and  illustrated in \fref{fig:tm_rel_uncertainty_comp}.
Here we restrict ourselves to the class of Fock coherent states.
The results of numerical calculations comparing the free and interacting cases for a range of parameters are given in \fref{fig:int_rel_uncertainty_comp}.
An immediate effect of the interactions is that the expectation values diverge at finite relational time, resulting in different ranges of the relational time coordinate.
These divergences can already be anticipated from the discussion below \eref{eq:int_classical_eff_friedmann_asymptotic}.
The key observation is that when interactions are present the relative uncertainties are not asymptotically constant but start growing as soon as the interactions become dominant.
The general statement that Fock coherent states become semiclassical at late times can therefore not be extended to this interacting case in an obvious way.
Again, this is also consistent with the expectation that mean-field methods break down in the interacting case when interactions begin to dominate over the quadratic Hamiltonian.

\section{Conclusions}
Our aim was to present a general perspective on the derivation of reliable effective Friedmann equations from given quantum dynamics of a \ac{gft} model, building on various recent developments in the derivation of effective cosmological dynamics from \ac{gft}.
Within this general perspective, the most important assumption was to restrict the \ac{gft} dynamics to those of a single field mode, i.e., fixed values of the representation labels in the Peter--Weyl decomposition.
This simplification of the full dynamics in which all modes would be present can be seen as the most important limitation of our work.
While there are arguments suggesting the dynamical emergence of a regime dominated by a single field mode in \ac{gft}, showing such an emergence in models of interest for four-dimensional quantum gravity remains an outstanding challenge.
On the other hand, we were able to derive effective cosmological dynamics without relying on a mean-field approximation, and in general no assumptions needed to be made on the initial state.

We first focussed on the case of dynamics defined by a free (quadratic) Hamiltonian.
Such a Hamiltonian can be of harmonic form or of `upside-down' harmonic form; the latter case in which the Hamiltonian is unbounded from below is most relevant to \ac{gft} cosmology, since it admits solutions expanding to infinity \cite{WEw19,GPW19}.
For this case, we recovered and extended results of \cite{AGW18} and \cite{BBC19} for the resulting effective Friedmann equations.
Generic solutions exhibit singularity resolution in the sense of a minimal non-zero value for the volume, and interpolate between a collapsing and an expanding branch which both at large volumes approach classical \ac{flrw} solutions.
These solutions depend on a parameter determined by the initial conditions (with no obvious classical analogue) which generates an asymmetry between the solution before and after the minimum for the volume.
The main new result in this part was a discussion of relative uncertainties in the two main physical observables---volume and energy---in three different classes of coherent states: Fock coherent states which have been used previously in \ac{gft} cosmology, and \acf{pg} and \acf{bg} coherent states of $\liealg{su}(1, 1)$.
We found that Fock coherent states approach a semiclassical regime at large volume, where relative uncertainties can be arbitrarily small, while PG states of interest here never reach such a regime.
The difference in our treatment compared with works such as \cite{EM12} was that we assumed the Fock space structure of \ac{gft} and only considered PG states that live in bosonic Fock representations.
In \ac{gft}, Fock coherent states are a good choice for initial conditions that become semiclassical at low curvature.
This part of our analysis could be broadened by going to more general types of coherent states for which only a Casimir condition and the saturation of uncertainty relations are assumed, as was done for $\liealg{su}(1, 1)$ in \cite{BT14}.

We then added a quartic interaction term to the Hamiltonian to extend the derivation of effective Friedmann equations for \ac{gft} models with polynomial interactions given in \cite{CPS16} to situations where no mean-field approximation is assumed.
As for the free case, one has a choice between a quartic term for which the Hamiltonian is bounded from below, which leads to a recollapse and cyclic cosmology \cite{CPS16}, or an interaction that admits solutions escaping to infinity, corresponding to a Universe expanding forever.
We focussed on the second case, the more common one in usual cosmology.
To understand the dynamics for this system, we first used a classical approximation in which one considers the basic dynamical variables as commuting phase space functions, with commutators replaced by Poisson brackets.
In this case we could derive an exact nonperturbative Friedmann equation; its unusual feature is the appearance of a square root involving the volume and energy on the right-hand side, preventing its straightforward interpretation in terms of effective perfect fluids.
Linearising this equation in the interaction constant leads to a Friedmann equation with an effective dust term, which would reproduce the mean-field result of \cite{CPS16}.
This linear correction only describes an intermediate regime in the expansion history, after which all orders become relevant.
We interpret this as signifying a failure of mean-field methods as soon as interactions become strong.
The asymptotic form of the effective Friedmann equation at late times would correspond to a matter component with equation of state $p=\frac{1}{2}\rho$ (instead of dust with $p=0$), or a modification of gravity on large scales in this model.
We then turned to the full quantum case in which one has to resort to a perturbative or numerical treatment.
For Fock coherent states, we find qualitatively similar results to the classical case: we derived a linearised correction to the effective Friedmann equation, and the full numerical solution quickly deviates from this regime as interactions become stronger.
We found numerical evidence that relative uncertainties in the volume start growing for Fock coherent states when the interactions become relevant, spoiling the property of these states to become semiclassical at late times that we observed for a quadratic Hamiltonian.

A main direction for future work will be lifting the assumption that only a single field mode contributes to the \ac{gft} dynamics.
Since the quadratic part of the full \ac{gft} Hamiltonian only couples pairs of modes, in the free case it would not be difficult to include additional modes into the analysis.
For this case one question would be whether some modes would always dominate asymptotically, as in the mean-field analysis of \cite{Gie16}.
When adding interaction terms however, we would expect the interaction of different modes to lead to a substantial modification of the effective cosmological dynamics away from the effectively free regime described by an \ac{lqc}-like bounce.
In particular this could apply to the recent proposal of \cite{GO18} for the generation of cosmological perturbations through quantum fluctuations in \ac{gft} cosmology.
In the long term, we would then also aim to bring the interacting \ac{gft} `toy' models studied in cosmological applications closer to candidate theories for full quantum gravity.

An entirely different but conceptually important direction would be to contrast the deparametrised framework used here, in which the scalar field $\phi$ is used as a clock from the beginning, with a covariant setting in which one is free to choose different clocks, following e.g. the ideas of \cite{Van18,Hoe18}.
We plan to investigate this in models which include multiple candidate matter clocks.

\section*{Acknowledgments}
We thank Martin Bojowald, Marco de Cesare, Daniele Oriti, Andreas Pithis and Edward Wilson-Ewing for helpful comments on an earlier version of the manuscript.
We also thank the referees for helpful suggestions for improvement.
The work of SG was funded by the Royal Society through a University Research Fellowship (UF160622) and a Research Grant for Research Fellows (RGF\textbackslash R1\textbackslash 180030).
AP was supported by the same Research Grant for Research Fellows awarded to SG.

\appendix
\section{Aspects of representation theory of \texorpdfstring{$\liealg{su(1, 1)}$}{su(1, 1)}}
\label{app:su11}

In this appendix we give a brief overview of the representation theory for $\liealg{su}(1, 1)$.

The Lie algebra $\liealg{su}(1,1)$ is defined by the following nonvanishing Lie brackets
\begin{eqnarray}
  &
  \commutator{
    \op{K}_{0}
  }{
    \op{K}_{\pm}
  }
  =
  \pm
  \op{K}_{\pm}
  \mathcomma
  \\
  &
  \commutator{
    \op{K}_{-}
  }{
    \op{K}_{+}
  }
  =
  2 \op{K}_{0}
  \mathperiod
\end{eqnarray}
The Casimir of this algebra is given by
\begin{equation}
  \label{eq:casimir}
  \op{C}
  =
  \op{K}_i
  \op{K}_j
  g^{ij}
  =
  (\op{K}_0)^2
  -
  \frac{1}{2}
  (
    \op{K}_{+}
    \op{K}_{-}
    +
    \op{K}_{-}
    \op{K}_{+}
  )
  \mathcomma
\end{equation}
where
\begin{equation}
  (g^{ij})
  =
  \left(
    \begin{array}{ccc}
      1 & 0 & 0 \\
      0 & 0 & -1/2 \\
      0 & -1/2 & 0
    \end{array}\right)
\end{equation}
is the Killing form which is used to raise indices.
Since $\liealg{su}(1,1)$ is noncompact its representations are infinite-dimensional.

We will start our discussion by considering representations in which the operator $\op{K}_{0}$ is diagonal and satisfies
\begin{equation}
  \op{K}_{0}
  \ket{\mu}
  =
  \mu
  \ket{\mu}
  \mathperiod
\end{equation}
The operators $\op{K}_{\pm}$ raise or lower the value of $\mu$ as can be seen from considering
\begin{equation}
  \op{K}_{0}
  \op{K}_{\pm}
  \ket{\mu}
  =
  (\mu \pm 1)
  \op{K}_{\pm}
  \ket{\mu}
  \mathperiod
\end{equation}
In particular this means that
\begin{equation}
  \op{K}_{\pm}
  \ket{\mu}
  =
  c_\pm(\mu)
  \ket{\mu \pm 1}
  \mathperiod
\end{equation}
By a straightforward computation one finds that
\begin{equation}
  \abs{
    c_\pm(\mu)
  }^2
  =
  \mu (\mu \pm 1) - c
  \mathcomma
\end{equation}
where $c$ is the eigenvalue of the Casimir operator \eref{eq:casimir}, i.e.\
\begin{equation}
  \op{C} \ket{\mu}
  =
  c \ket{\mu}
  \mathperiod
\end{equation}
The representations can be characterised by studying if the coefficients $c_\pm(\mu)$ vanish for some value of $\mu$.
Since we are mostly interested in interpreting the operator $\op{K}_{0}$ as a (shifted and rescaled) number operator, we will restrict ourselves to the case where the action of $\op{K}_{-}$ annihilates the `initial' state after some iterations.
In particular, if one assumes that $c_{-}(k) = 0$ one immediately gets
\begin{equation}
  c
  =
  k (k - 1)
  \mathperiod
\end{equation}

The representations thus available are characterised by a positive real number $k$.
We will denote the state that is annihilated by $\op{K}_{-}$ as $\ket{k, 0}$ and the rest of the representation can be generated by acting iteratively with the operator $\op{K}_{+}$.
We define the states to be normalised as
\begin{equation}
  \label{eq:state}
  \ket{k, m}
  =
  \sqrt{
    \frac{
      \Gamma(2k)
    }{
      \Gamma(2k + m) m!
    }
  }
  (\op{K}_{+}{})^m
  \ket{k, 0}
  \mathperiod
\end{equation}
These states satisfy
\begin{eqnarray}
  &
  \op{C}
  \ket{k, m}
  =
  k (k-1)
  \ket{k, m}
  \mathcomma
  \\
  &
  \op{K}_{0}
  \ket{k, m}
  =
  (k + m)
  \ket{k, m}
  \mathperiod
\end{eqnarray}

To understand the normalising coefficient appearing in \eref{eq:state} it is helpful to write the action of the operators $\op{K}_{\pm}$ similarly as before,
\begin{equation}
  \op{K}_\pm
  \ket{k, m}
  =
  c_\pm(k, m)
  \ket{k, m \pm 1}
  \mathperiod
\end{equation}
From this one readily gets
\begin{eqnarray}
  &
  \abs{
    c_+(k, m)
  }^2
  =
  (2k + m)(m + 1)
  \mathcomma
  \\
  &
  \abs{
    c_-(k, m)
  }^2
  =
  (2k + m - 1)m
  \mathperiod
\end{eqnarray}

\subsection{Realisation as bosonic operators}
\label{sec:su11_bosonic_realisation}

The Lie algebra $\liealg{su}(1,1)$ can be realised by bosonic operators $\op{a}$, $\hermconj{\op{a}}$ satisfying the standard commutation relations $\commutator{\op{a}}{\hermconj{\op{a}}} = \op{I}$.
A suitable identification is as follows
\begin{equation}
  \op{K}_{0} = \frac{1}{4} ( \op{a} \hermconj{\op{a}} + \hermconj{\op{a}} \op{a} )
  \mathcomma
  \qquad
  \op{K}_{+} = \frac{1}{2} \hermconj{\op{a}}{}^2
  \mathcomma
  \qquad
  \op{K}_{-} = \frac{1}{2} \op{a}^2
  \mathperiod
\end{equation}
The Casimir then turns out to be identically
\begin{equation}
  \op{C} = - \frac{3}{16} \op{I}
  \mathperiod
\end{equation}
Recalling that for the discrete series $\op{C} = k(k-1) I$, one finds that either $k = 1/4$ or $k = 3/4$.
By considering
\begin{equation}
  \op{K}_{0}
  \ket{k, m}
  =
  (k + m)
  \ket{k, m}
\end{equation}
one can see that the states $\ket{k, m}$ correspond to the standard bosonic states $\ket{n} = (n!)^{-1/2} \hermconj{\op{a}}{}^n \ket{0}$ in the following way
\begin{equation}
  \ket{1/4, 2n}
  \equiv
  \ket{2n}
  \mathcomma
  \qquad
  \ket{3/4, 2n + 1}
  \equiv
  \ket{2 n + 1}
  \mathperiod
\end{equation}

\subsection{Coherent states}
\label{sec:coherent_states}

For the Lie algebra $\liealg{su}(1,1)$ there are several distinct notions of coherent states.
These arise naturally when trying to generalise the different ways one can construct coherent states in the well-known case of the harmonic oscillator.
There, for instance one might view coherent states as the eigenvectors of the annihilation operator or to be generated by the action of a specific operator acting on ground state.
In the case of the harmonic oscillator these notions coincide.
However, in the more general setting these notions lead to different coherent states.
We will discuss two notions of coherent states.
Firstly, we will discuss coherent states generated by acting on the ground state by some specific operator which are commonly referred to as \acf{pg} coherent states since they were discussed in detail in \cite{Per86,Per72,Gil72}.
Secondly, coherent states can be characterised as being eigenstates of the lowering operator.
For $\liealg{su}(1,1)$ these states are referred to as \acf{bg} states \cite{BG71}.

A general discussion of coherent states can be found in the textbooks \cite{Per86,CR12}.

\subsection{Perelomov--Gilmore coherent states}

The operator which generates the coherent states from the coherent states is given by
\begin{equation}
  \op{S}(\xi)
  =
  \exp(
    \xi \op{K}_{+}
    -
    \compconj{\xi}
    \op{K}_{-}
  )
  \mathperiod
\end{equation}
This operator can be written in the `normal-ordered' form \cite{Per86}
\begin{equation}
  \label{eq:displacement_alt}
  \op{T}(\zeta)
  =
  \expe^{\zeta \op{K}_{+}}
  \expe^{\eta \op{K}_{0}}
  \expe^{- \compconj{\zeta} \op{K}_{-}}
  \mathcomma
\end{equation}
where $\zeta$ is defined as
\begin{equation}
  \zeta
  =
  \frac{\xi}{\abs{\xi}}
  \tanh \abs{\xi}
  \mathcomma
  \quad
  \eta
  =
  \ln ( 1 - \tanh^2 \abs{\xi} )
  =
  \ln ( 1 - \abs{\zeta}^2 )
  \mathcomma
\end{equation}
These relations can also be expressed in the condensed form
\begin{equation}
  \op{T}(\zeta)
  =
  \op{S}\left(
    \frac{\zeta}{\abs{\zeta}}
    \artanh\abs{\zeta}
  \right)
  \mathperiod
\end{equation}

The product of two such $\op{T}$ operators can be explicitly given as
\begin{equation}
  \label{eq:displacement_product}
  \op{T}(\zeta_1)
  \op{T}(\zeta_2)
  =
  \exp\left(
    \ln\left(
      \frac{
        1 + \zeta_1 \compconj{\zeta}_2
      }{
        1 + \compconj{\zeta}_1 \zeta_2
      }
    \right)
    \op{K}_0
  \right)
  \op{T}(\zeta_3)
  \mathcomma
  \qquad
  \zeta_3
  =
  \frac{
    \zeta_1 + \zeta_2
  }{
    1 + \zeta_1 \compconj{\zeta}_2
  }
\end{equation}
or with the opposite ordering
\begin{equation}
  \op{T}(\zeta_1)
  \op{T}(\zeta_2)
  =
  \op{T}(\zeta_3)
  \exp\left(
    \ln\left(
      \frac{
        1 + \zeta_1 \compconj{\zeta}_2
      }{
        1 + \compconj{\zeta}_1 \zeta_2
      }
    \right)
    \op{K}_0
  \right)
  \mathcomma
  \qquad
  \zeta_3
  =
  \frac{
    \zeta_1 + \zeta_2
  }{
    1 + \compconj{\zeta}_1 \zeta_2
  }
  \mathperiod
\end{equation}

Using the `normal-ordered' form \eref{eq:displacement_alt} of the displacement operator one finds acting on the `ground state'
\begin{equation}
  \ket{\zeta, k}
  =
  \op{T}(\zeta) \ket{k, 0}
  =
  (1 - \abs{\zeta}^2)^k
  \sum_{m = 0}^\infty
  \sqrt{
    \frac{
      \Gamma(2k + m)
    }{
      \Gamma(2k)
      m!
    }
  }
  \zeta^m
  \ket{k, m}
  \mathperiod
\end{equation}

Expectation values of `anti-normal-ordered' operators can be given in a closed
form \cite{Lis92}
\begin{eqnarray}
  \label{eq:pg_coherent_expval}
  &
  \bra{\zeta, k}
  (\op{K}_{-})^p
  (\op{K}_{0})^q
  (\op{K}_{+})^r
  \ket{\zeta, k}
  \\
  &\quad=
  \frac{
    (1 - \abs{\zeta}^2)^{2k}
  }{
    \Gamma(2k)
  }
  \zeta^{p - r}
  \\
  &\qquad
  \times
  \sum_{m=0}^\infty
  \frac{
    \abs{\zeta}^{2m}
  }{m!}
  \frac{
    \Gamma(2k + m + p)
    \Gamma(m + p + 1)
  }{
    \Gamma(m + p - r + 1)
  }
  (k + m + p)^q
  \mathperiod
\end{eqnarray}

The \ac{pg} coherent states are eigenstates of the operator $v^i \op{K}_i$, where \cite{Sch16}
\begin{equation}
  (v_0, v_+, v_-)
  =
  \frac{1}{1 - \abs{\zeta}^2}
  (
    1 + \abs{\zeta}^2,
    2 \compconj{\zeta},
    2 \zeta
  )
\end{equation}
with eigenvalues $k$, i.e.
\begin{equation}
  v^i \op{K}_i
  \ket{\zeta, k}
  =
  g^{ij}
  v_i \op{K}_j
  \ket{\zeta, k}
  =
  k
  \ket{\zeta, k}
  \mathperiod
\end{equation}

As an aside we remark that it was pointed out in \cite{EM12} that the \ac{pg} coherent states saturate the uncertainty relations
\begin{equation}
  \covariance{\op{A}}{\op{A}}
  \covariance{\op{B}}{\op{B}}
  \geq
  \covariance{\op{A}}{\op{B}}^2
  +
  \expval{
    \frac{1}{2\imagi}
    \commutator{\op{A}}{\op{B}}
  }^2
  \mathcomma
\end{equation}
where we recall that the covariance $\covariance{\op{A}}{\op{B}}$ is defined as
\begin{equation}
  \covariance{\op{A}}{\op{B}}
  =
  \expval{
    \frac{1}{2}
    (\op{A}\op{B} + \op{B}\op{A})
  }
  -
  \expval{\op{A}}
  \expval{\op{B}}
  \mathperiod
\end{equation}

\subsection{Barut--Girardello coherent states}

As discussed above, an alternative notion of coherent state can be that the coherent state is an eigenstate of the lowering operator.
This is the definition used in \cite{BG71}.

The coherent states satisfy
\begin{equation}
  \op{K}_{-}
  \ket{\chi, k}
  =
  \chi
  \ket{\chi, k}
  \mathperiod
\end{equation}
They are given by
\begin{eqnarray}
  &
  \ket{\chi, k}
  =
  N(\chi, k)
  \sum_{m=0}^\infty
  \frac{
    \chi^m
  }{
    \sqrt{
      m!
      \Gamma(2k + m)
    }
  }
  \ket{k, m}
  \mathcomma
  \\
  &
  N(\chi, k)
  =
  \sqrt{
    \frac{
      \abs{\chi}^{2k-1}
    }{
      I_{2k - 1}(2 \abs{\chi})
    }
  }
  \mathcomma
\end{eqnarray}
where $I_\alpha(x)$ is the modified Bessel function of the first kind.

For `normal-ordered' products of $\liealg{su}(1,1)$ elements the expectation value for \ac{bg} coherent states is given by
\begin{equation}
  \fl
  \bra{\chi, k}
  (\op{K}_+)^p
  (\op{K}_0)^q
  (\op{K}_-)^r
  \ket{\chi, k}
  =
  \frac{
    \abs{\chi}^{2k}
  }{
    I_{2k-1}(2 \abs{\chi})
  }
  \compconj{\chi}^p
  \chi^r
  \sum_{m=0}^\infty
  \frac{
    \abs{\chi}^{2m}
  }{
    m!
    \Gamma(2k + m)
  }
  (k + m)^q
  \mathperiod
\end{equation}

\subsection{Fock coherent states}

As explained in \sref{sec:su11_bosonic_realisation} $\liealg{su}(1, 1)$ can be realised in terms of bosonic creation and annihilation operators.
For this Fock space representation there are the well-known coherent states of the harmonic oscillator which can be characterised as being eigenstates of the annihilation operator, i.e.\ $\op{a} \ket{\sigma} = \sigma \ket{\sigma}$.
For `normal-ordered' operators the expectation value for Fock coherent states is given by
\begin{eqnarray}
  \bra{\sigma}
  (\op{K}_+)^p
  (\op{K}_0)^q
  (\op{K}_-)^r
  \ket{\sigma}
  =
  \frac{1}{
    2^{p+q+r}
  }
  \compconj{\sigma}^p
  \sigma^r
  \sum_{m=0}^{q}
  {q \choose m}
  \frac{1}{2^{q-m}}
  \bra{\sigma}
    \op{N}^m
  \ket{\sigma}
  \mathcomma
  \\
  \bra{\sigma}
    \op{N}^m
  \ket{\sigma}
  =
  \sum_{n=0}^m
  S(m, n)
  \abs{\sigma}^{2n}
  \mathcomma
\end{eqnarray}
where $S(m, n)$ are the Stirling numbers of the second kind.

\subsection{Central extension attempt}\label{sec:central_extension_attempt}

Due to the fact that the operator $\op{K}_0$ can be interpreted as a shifted and rescaled version of the bosonic number operator, one might be tempted to centrally extend $\liealg{su}(1,1)$ by defining the following relation
\begin{equation}
  \commutator{
    \op{K}_-
  }{
    \op{K}_+
  }
  =
  2 \op{K}_0
  +
  c \op{I}
  \mathcomma
\end{equation}
where $\op{I}$ is a central element.
We denote the centrally extended algebra by $\liealg{su}(1,1)_c$.
The Casimir of this algebra is given by\footnote{From a Lie algebraic perspective one might be inclined to add some term proportional to the central element $\op{I}$.
  Such an additional term changes only the eigenvalues of the Casimir and the rest of the representation theory discussed here does not change.
}
\begin{equation}
  \op{C}
  =
  (\op{K}_0)^2
  +
  c \op{K}_0
  -
  \frac{1}{2}
  \left(
    \op{K}_+
    \op{K}_-
    +
    \op{K}_-
    \op{K}_+
  \right)
  \mathperiod
\end{equation}

Performing the same steps as above one arrives at the discrete rising series
\begin{eqnarray}
  &
  \op{K}_0
  \ket{k, m}
  =
  (k + m)
  \ket{k, m}
  \mathcomma
  \\
  &
  \op{C}
  \ket{k, m}
  =
  k(k + c -1)
  \ket{k, m}
  \mathcomma
  \\
  &
  \ket{k, m}
  =
  \sqrt{
    \frac{
      \Gamma(2k + c)
    }{
      m!
      \Gamma(2k + m + c)
    }
  }
  (
    \op{K}_+
  )^m
  \ket{k, 0}
  \mathcomma
\end{eqnarray}
where $k > - c/2$ and $m$ is a nonnegative integer.
From these formulas it becomes evident that these representations can be obtained from the case without central extension by the replacements $k \rightarrow k + c/2$, $\op{K}_0 \rightarrow \op{K}_0 - \frac{c}{2} \op{I}$ and $\op{C} \rightarrow \op{C} - \frac{c}{2}(\frac{c}{2} - 1)\op{I}$.
 
\section*{References}
\providecommand{\newblock}{}

\end{document}